\documentclass[]{raa}            
\usepackage{graphicx,times}
\usepackage{natbib}

\usepackage{amsmath}
\usepackage{amssymb}
\usepackage{multirow}
\usepackage{xcolor}

\usepackage{setspace}
\providecommand{\bm}[1]{\mbox{\boldmath$#1$\unboldmath}}

\begin{document}

   \title{Rotating massive strangeon stars and X-ray plateau of short GRBs}

\volnopage{ {\bf 2023} Vol.\ {\bf X} No. {\bf XX}, 000--000}
   \setcounter{page}{1}

   \author{Xi-Yan Yang\inst{1,2}, Xiao-Yu Lai\inst{2}, Wei-Wei Tan\inst{2,3}, Ren-Xin Xu\inst{4,5}
   }

   \institute{School of Physics and Optoelectronic Engineering, Yangtze University, Jingzhou 434023, China\\
        \and
        Research Center for Astronomy, Hubei University of Education, Wuhan 430205, China; {\it laixy@hue.edu.cn}\\
        \and
             Key Laboratory of Quark and Lepton Physics (Central China Normal University), Ministry of Education, Wuhan 430079, China\\
     \and
     School of Physics, Peking University, Beijing 100871, China\\
     \and
     Kavli Institute for Astronomy and Astrophysics, Peking University, Beijing 100871, China\\
      \vs \no
   {\small Received [year] [month] [day]; accepted [year] [month] [day]}
}

\abstract{Strangeon stars, which are proposed to describe the nature of pulsar-like compact stars, have passed various observational tests.
The maximum mass of a non-rotating strangeon star could be high, which implies that the remnants of binary strangeon star mergers could even be long-lived massive strangeon stars.
We study rigidly rotating strangeon stars in the slowly rotating approximation, using the Lennard-Jones model for the equation of state.
Rotation can significantly increase the maximum mass of strangeon stars with unchanged baryon numbers, enlarging the mass-range of long-lived strangeon stars.
During spin-down after merger, the decrease of radius of the remnant will lead to the release of gravitational energy.
Taking into account the efficiency of converting the gravitational energy luminosity to the observed X-ray luminosity, we find that the gravitational energy could provide an alternative energy source for the plateau emission of X-ray afterglow.
The fitting results of X-ray plateau emission of some short gamma-ray bursts suggest that the magnetic dipole field strength of the remnants can be much smaller than that of expected when the plateau emission is powered only by spin-down luminosity of magnetars.
 }

   \authorrunning{X.-Y. Yang et al. }            
   \titlerunning{Rotating massive SSs and X-ray plateau of SGRBs}  
   \maketitle

%

\section{Introduction}
\label{sec:introduction}

Pulsar-like compact stars, among which the well-known ones are radio pulsars\footnote{They are strictly called ``pulsar-like compact stars'' instead of ``pulsars'' because some of them do not manifest themselves as radio pulsars. In most case if it will not cause misunderstandings, they are still called ``pulsars'' for short.}, are born in supernova explosions.
Although abundant observational data of pulsars have been accumulated, their interior structure is still a controversial topic at present.
The typical density of a pulsar is slightly larger than that of nuclear matter on average, so the equation of state (EoS) of pulsars essentially depends on the behavior of quantum chromodynamics (QCD) at low-energy scales, which is still a challenge for us to understand.
Some basic problems still remain. Have quarks been de-confined there? Does strangeness play an essential role?
Based on different points of view, a variety of models for pulsars have been speculated, such as neutron stars (NSs) and quark stars (QSs).
A strangeon star (SS) model was proposed originally by~\citet{Xu2003} and has been studied for twenty years, based on both observational and theoretical points of view (see~\cite{LXX2023} and references therein).
Briefly speaking, at realistic densities inside pulsars, the energy scale is much higher than the mass difference between strange ($s$) and up/down ($u/d$) quarks, while is additionally not high enough to justify the validity of perturbative QCD.
The net strangeness could emerge due to the weak interaction, and $u$, $d$ and $s$ quarks would tend to be of the same amount in the light-flavor symmetry.
At the same time, the non-perturbative effects of QCD could be significant, similar to the case of the strong interaction in nuclei.
It is then conjectured that strangeons\footnote{Strangeons were previously called ``strange quark-clusters'' in some earlier papers. The strong coupling between quarks may group quarks into clusters which are composed of several quarks.}, which can be understood as ``nucleons with strangeness'' or ``strange nucleons'', could be the building blocks of dense matter in pulsars~\citep{Xu2003,XG2016}.

Compact stars composed totally of strangeons are called strongeon stars (SSs).
SSs are different from both NSs and strange quark stars (SQSs).
Different from baryons inside NSs, strangeons are of three-flavored, and the number of constituent quarks ($N_{\rm q}$) of a strangeon can be larger than three.
For instance, strangeons with $N_{\rm q}=18$ are similar to the so called quark-alpha~\citep{Michel1988} which are be completely asymmetric in spin, flavor and color space.
Moreover, NSs are gravitational bound and have crusts composed of normal nuclei, while SSs are self-bound and have almost the same composition from the center to the surface.
SSs are also different from SQSs, and the main difference is that SSs are self-bound by the residual interaction between strangeons, while SQSs are self-bound by bag-like confinement.

The SS model has passed various observational tests.
It predicted high-mass pulsars (possibly even larger than $3M_\odot$)~\citep{LX2009a,LX2009b} before the formal discovery of massive pulsars with $M>2M_\odot$.
In addition, the strangeon matter surface could naturally explain the pulsar magnetospheric activity~\citep{Xu1999ApJL} as well as the subpulse-drifting of radio pulsars~\citep{Lu2019}.
Pulsar glitch~\citep{Zhou2004,Zhou2014,Lai2018MN} and glitch recovery~\citep{Lai2023MN} can also be explained under the framework of starquake of SSs.
The plasma atmosphere of SSs can reproduce the Optical/UV excess observed in X-ray dim isolated neutron stars~\citep{Kaplan2011,Wang2017APJ}.
The tidal deformability of merging binary SSs is consistent with the results of gravitational wave event GW170817~\citep{Lai2018RAA,LZX2019}.

Rotation will affect the structure of pulsars, and related astrophysical consequences are worth exploring to provide tests for EoS models.
A perturbative approach describing distorted NSs with uniform and slow rotation to the second order of angular frequency $\Omega$ was given by~\citet{Hartle1967} and~\citet{Hartle1968}, and was developed to the third order of $\Omega$ by~\citet{Hartle1973} for calculating the variations of moments of inertia.
Slow rotation means that $\Omega$ of a star with mass $M$ and radius $R$ is much smaller than a critical frequency, $\Omega\ll \Omega_{\rm c} = \sqrt{GM/R^3}$, in which case the rotating configuration can be considered as a perturbation on a non-rotating one of the same central density.
It has been shown that this perturbative approach can be applied with great accuracy for most observed NSs, even for most millisecond pulsars~\citep{Berti2004,Benhar2005,Berti2005}.
\citet{Gao2022} has provided detailed calculations about the structure of slowly rotating SSs, using the EoS of the Lennard-Jones model, and derived the moments of inertia, the quadrupole moments, the eccentricities, changes in the gravitational and baryonic masses, and universal relations between some of these quantities.

The effect of rotation on the stability is crucial for the fate of the products of binary NS/SS mergers.
The maximum mass of non-rotational NSs/SSs, denoted by $M_{\rm TOV}$, can be derived by solving the Tolman-Oppenheimer-Volkoff (TOV) equations~\citep{OV1939} for a given EoS.
According to the widely used definition, the rotating NS/SS that is stablized by differential rotation is {\it hypermassive}, the one that is stablized by rigid rotation is {\it supramassive}, and the one that is stable without rotation is {\it stable} or {\it long-lived}.
The maximum mass of rotating NSs/SSs, denoted by $M_{\rm max}$, will increase with the angular frequency $\Omega$, which also depends on EoSs.
Considering that the EoSs should satisfy the constraints of both the existence of two-solar-mass pulsars and the tidal deformability of GW 170817~\citep{GW170817}, we use the EoS of the Lennard-Jones model~\citep{LX2009b} (which has also been used by~\citet{Gao2022}) for SSs and the EoS of the AP4 model~\citep{AP1997} for NSs.

The fate of the product of a merger event should be determined in a constant baryon number.
We will explicitly show the increases of $M_{\rm max}$ with $\Omega$ along the lines of constant baryonic mass, with the results indicating that the increases of $M_{\rm max}$ are more pronounced for SSs than those for NSs.
Combining with the previous conclusion that $M_{\rm TOV}$ of SSs can be larger than $2.5 M_\odot$ in a wide range of parameter space~\citep{LX2009a,LX2009b,LGX2013,GLX2014}, we can infer that if pulsar-like compact stars are actually SSs, the remnants of binary strangeon star mergers are very likely to be long-lived massive SSs.
The long-lived SSs as the remnants of binary strangeon star mergers could have interesting observational consequences, e.g. they could reproduce the light curves observed in kilonova~\citep{Lai2018RAA,Lai2021RAA}.
As will be demonstrated in this paper, the long-lived massive SSs could also provide large gravitational energy enough to explain the X-ray afterglow of short gamma-ray bursts (GRBs).

Short gamma-ray bursts (SGRBs) are generally believed to originate from the binary NS mergers~\citep{Eichler1989} or the binary NS-BH (black hole) mergers~\citep{Paczynski1991}.
Among them, the ones with afterglow phase are often interpreted as originating from binary NS mergers, where the afterglow emission is widely accepted as being powered by the millisecond magnetars~\citep[e.g.,][]{Dai1998,Zhang2001}.
The electromagnetic dipolar radiation of the postmerger magnetars could explain the X-ray flares following SGRBs~\citep{DaiZG2006,GaoWH2006}, and the lightcurves produced by magnetar spindown winds could explain the X-ray plateaus at observed luminosities of SGRBs~\citep{Strang2019,Strang2021}.
The gravitational bursts due to magnetar wind dissipation of the millisecond magnetars left behind binary neutron star mergers~\citep{ZhangB2013} and the associated multi-wavelength afterglows~\citep{GaoH2013} have been investigated. Moreover, some plateaus with long durations are suggested to be powered by the nascent SQSs~\citep{YuYW2009}, which could be supported by observations on the break time of internal X-ray plateaus in SGRBs~\citep{LiA2016}.
The X-ray plateaus in the afterglow of GRBs observed by {\it Swift} satellite have been explained by the magnetar central engines, where the dipole field with strength larger than $10^{15}$ G is required~\citep{Rowlinson2013,Stratta2018}.
In this paper, we will investigate the gravitational energy released by long-lived massive SSs and the implication on the X-ray afterglow of SGRBs.

If pulsar-like compact stars are actually SSs instead of NSs, i.e. the binary neutron star mergers are actually binary SS mergers, then SGRBs with the plateaus in the X-ray afterglow phase originate from binary SS mergers.
It is worth noting that, the strong magnetic field may not be necessary for the scenario of SSs, e.g. the elastic and gravitational energy released of SSs can explain the AXPs/SGRs (anomalous X-ray pulsars/soft gamma repeaters) associated with glitches~\citep{Zhou2014} and the precursor emission of SGRBs~\citep{Zhou2023}.
The latent heat released in solidification of the strangeon star was proposed as the energy injection into the X-ray plateau of GRB afterglow~\citep{DLX2011}, and this idea is supported by the X-ray light curve of GRB 170714A whose two plateaus can be interpreted as being powered respectively by the latent heat and the spin-down of a massive strangeon star which is the remnant of the binary star merger~\citep{Hou2018}.
For the sake of simplicity, it was assumed that the latent heat released in solidification was released as the blackbody radiation to power the relativistic jet of GRBs; however, more realistic and detailed mechanism should be taken into account.

If pulsar are actually SSs instead of NSs, it is worth exploring explicitly that how the plateaus in the X-ray afterglows of SGRBs originate from binary SS mergers. In this paper, we will consider the contribution of gravitational energy to the X-ray light curves of GRBs from binary star mergers.
The remnants of binary SS mergers will undergo spin-down due to energy loss.
The gravitational mass $M$ and radius $R$ will reduce as the angular velocity $\Omega$ reduces under the same baryonic mass $M_{\rm b}$, so the gravitational energy will be released as an isolate star is spinning-down.
Because the remnants of binary SS mergers would be long-lived massive SSs, we will study the spin-down process of massive SSs and investigate whether the gravitational energy could provide the energy injection to the X-ray plateau in the afterglow of SGRBs.
As we will demonstrate in this paper, the shrinkage of the star would lead to oscillations and turbulence, which would convert the gravitational energy into kinematic energy and finally injected into the GRB fireball.

Assuming that the spin-down is due to magnetic dipolar radiation, we can derive the luminosity of gravitational energy releasing, then the X-ray luminosity can be derived by taking into account the efficiency of converting the gravitational energy to the observed X-ray luminosity.
From~\citet{Stratta2018} which interpreted the GRBs presenting X-Ray afterglow plateaus as of the magnetar origin, we choose some SGRBs with obvious plateaus in {\it Swift} GRB sample, and fit their X-ray afterglow data using the MCMC (Markov Chain Monte Carlo) method.
Our fitting results will be compared to that in~\citet{Stratta2018}.
The values of $\chi^2/\rm{d.o.f}$ in our scenario are not much larger (and in some cases are even smaller) than that in~\citet{Stratta2018}.
In addition, the results show that the magnetic dipole field strength of the remnants can be much smaller than that of expected when the plateau emission is powered only by spin-down luminosity of magnetars.

This paper is organized as follows.
After the introduction about the Lennard-Jones model of SSs in \S\ref{subsec:static}, we demonstrate how to calculate the structure of slowly rotating SSs in the Hartle-Thorne approximation in \S\ref{subsec:rotation}, and show the results for maximum mass and spherical stretching under rotation.
Based on the luminosity of gravitational energy during spin-down derived in \S\ref{subsec:Lgrav}, we investigate in \S\ref{subsec:fit} whether the gravitational energy release during spin-down could provide enough energy injection for the plateau emission of X-ray afterglow of SGRBs.
Conclusions and discussions are made in \S\ref{sec:conclusion}.

\section{Strangeon stars in slow rotation approximation}
\label{sec:rotation}

\subsection{Static strangeon stars}
\label{subsec:static}

We choose the Lennard-Jones model to describe the EoS of SSs~\citep{LX2009b} because it can well characterizes the non-relativistic nature and the strong-repulsive interactions at short distances, and the allowed parameter space to satisfy the constraints by both the existence of two-solar-mass pulsars and the tidal deformability of GW 170817 is large~\citep{LZX2019}.
The potential between two strangeons is
\begin{equation}
u(r)=4U_0\left[\left(\frac{r_0}{r}\right)^{12}-\left(\frac{r_0}{r}\right)^6\right],
\end{equation}
where $U_0$ is the depth of the potential and $r_0$ is the range of interaction.
The total energy density includes the densities of the potential energy, the lattice vibration energy and the baryonic mass-energy.
The lattice energy density is negligible compared to the other two energy densities, so the total energy density is
\begin{equation}
\rho c^2=2U_0\left(6.2r_0^{12}n^5-8.4r_0^6n^3\right)+nmc^2,
\end{equation}
and the pressure is
\begin{equation}
P=4U_0\left(12.4r_0^{12}n^5-8.4r_0^6n^3\right),
\end{equation}
where $n$ is the number density of strangeons, $m$ is the mass of each strangeon, and the simple-cubic lattice structure is assumed.
If the number of quarks in each strangeon is $N_{\rm q}$, then $m\simeq N_{\rm q}\cdot 300$ MeV.
The parameter $r_0$ is related to the baryon number density on the surface $n_{\rm b,s}$ where the pressure vanishes.

As in~\citet{Gao2022}, we choose the EoS with parameters $n_{\rm b,s}=0.36$ fm$^{-3}$, $U_0=30$ MeV (denoted by LX3630) and $N_{\rm q}=18$, because it satisfies the constraint from the measurement of the moment of inertia of PSR J0737-3039A~\citep{Hu2020}.
Given the EoS, the structure of non-rotating SSs is governed by the TOV equations~\citep{OV1939}.
The gravitational mass is $M=4\pi\int_0^R\rho r^2 {\rm d}r$ and baryonic mass is $M_{\rm b}=4\pi \int_0^R \rho_{\rm b}e^{\lambda} r^2 {\rm d}r$, where $\rho_{\rm b}=nm$ is the baryon density, $e^{\lambda}=1/\sqrt{1-2Gm/(rc^2)}$ and $m$ is the gravitational mass enclosed in radius $r$.

\subsection{Strangeon stars under slow rotation}
\label{subsec:rotation}

Given a central density $\rho_{\rm c}$ and the EoS, the structure derived by the TOV equation is the static and spherical background, based on which the gravitational mass $M$, the radius $R$ at equator, and the baryonic mass $M_{\rm b}$ of a rigidly rotating star in slow rotation approximation can be derived by adding corrections to the second order of $\Omega$.
This procedure was first formulated by~\citet{Hartle1967} and~\citet{Hartle1968} for rotating NSs.
The structure of rotating SSs was given in details by~\citet{Gao2022}, including the corrections induced by the match conditions on the surface.
Here we adopt the same procedure as that in~\citet{Gao2022} to show the evolution along given values of $M_{\rm b}$.
The values of $M$, $M_{\rm b}$ and $R$ are calculated to the spherical terms in the second order of $\Omega$, so only spherical deformations are considered.
The moment of inertia $I$ in \S\ref{subsec:Lgrav} is calculated by taking into account the corrections to the third order of $\Omega$ to the angular moment $J$.

The calculation can be proceeded as follows, whose details can be found in~\citet{Gao2022}.
(i) Choose a central density $\rho_{\rm c}$ to calculate the structure of a non-rotating configuration.
(ii) The gravitational mass $M$ and the radius $R$ of a rotating SS with the angular frequency $\Omega_{\rm c}=\sqrt{GM/R^3}$ is derived by adding the perturbations to a non-rotating configuration under the same $\rho_{\rm c}$, taking into account the matching condition at the surface.
(iii) The values of $M$ and $R$ of the configuration with a different angular frequency $\Omega (<\Omega_{\rm c})$ and the same central density can be obtained by multiplying the perturbations by the rescaling factor $(\Omega/\Omega_{\rm c})^2$.
Changing the values of $\rho_{\rm c}$ gives the $M-R$ curve. The $M-R$ curve with another value of $\Omega (<\Omega_{\rm c})$ can be derived from the same procedure.
(iv) By connecting the same value of $M_{\rm b}$ on each $M-R$ curve we can get a constant-$M_{\rm b}$ line.

\subsubsection{Evolution under constant baryonic mass}
\label{subsubsec:Mb}

\begin{figure}[h!!!]
\centering
\includegraphics[width=9.0cm, angle=0]{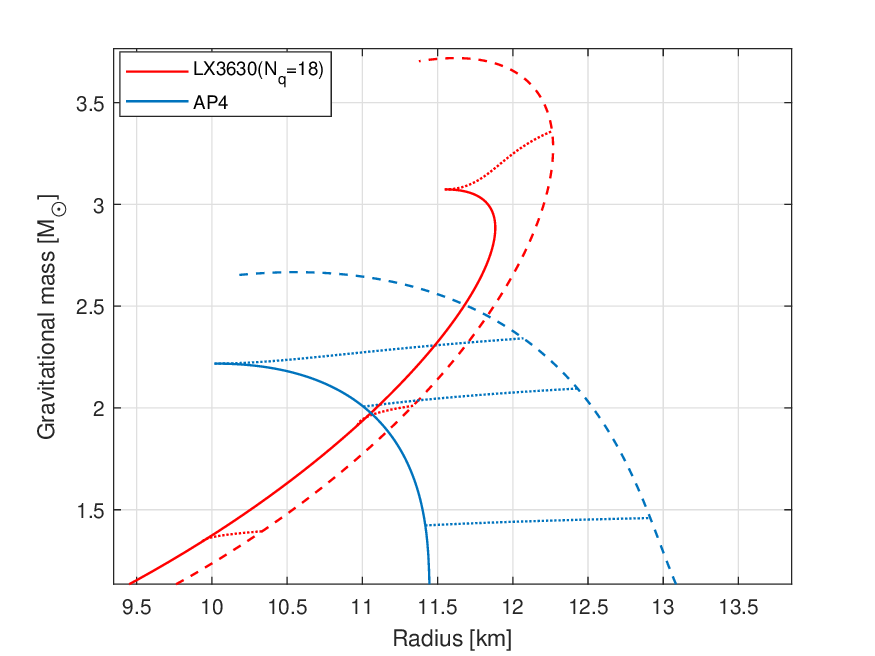}
\begin{minipage}[]{85mm}
\caption{$M-R$ curves for SSs (red curves) and NSs (blue curves), where solid lines represent non-rotating configurations, and dashed lines represent rotating configurations with the angular frequency $\Omega_{\rm c}$. Sequences of constant-$M_{\rm b}$ are denoted by dotted lines with $M_{\rm b} = M_{\rm b,max}^{\rm stable}$, $2.4 M_{\odot}$, and $1.6 M_{\odot}$, where $M_{\rm b,max}^{\rm stable}|_{\rm SS}\simeq 4.4 M_{\odot}$ and $M_{\rm b,max}^{\rm stable}|_{\rm NS}\simeq 2.7 M_{\odot}$.}
\end{minipage}
\label{fig_rhocb}
\end{figure}

An isolated star has an unchanged baryonic mass during spin-down.
For a given EoS, the stable configuration with $M=M_{\rm TOV}$ has the baryonic mass $M_{\rm b,max}^{\rm stable}$, and the sequences with baryonic mass $M_{\rm b}\leqslant M_{\rm b,max}^{\rm stable}$ will evolve to the stable configurations with the unchanged baryonic mass as they spin down.
We plot the gravitational mass and radius curves of strangeon stars in Fig.\ref{fig_rhocb}, under EoS LX3630 with $N_{\rm q}=18$, in which case $M_{\rm b,max}^{\rm stable}|_{\rm SS}\simeq 4.4 M_{\odot}$.
The sequences of constant-$M_{\rm b}$ are denoted by the red dotted lines with $M_{\rm b} = 4.4 M_{\odot}$, $2.4 M_{\odot}$, and $1.6 M_{\odot}$.
For comparison, we also plot the result of NSs under EoS AP4~\citep{AP1997}, in which case $M_{\rm b,max}^{\rm stable}|_{\rm NS}\simeq 2.7 M_{\odot}$, and the sequences of constant-$M_{\rm b}$ are denoted by the blue dotted lines with $M_{\rm b} = 2.7 M_\odot$, $2.4 M_{\odot}$, and $1.6 M_{\odot}$.
Solid lines represent non-rotating configurations, and dashed lines represent rotating configurations with the critical angular frequency $\Omega_{\rm c}$.

Our results show that the increase of gravitational mass due to rotation will be larger for larger mass, and the increases of $M_{\rm max}$ by rigid rotation for SSs are more pronounced than that for NSs.
The $M_{\rm max}$ value for SSs is roughly 9.3\% higher than $M_{\rm TOV}$ along the constant baryonic mass lines, and for the case of NSs the result is roughly 5.6\%.
If only compared with the maximum values of the $M-R$ curves, both the results are roughly 20\% for SSs and NSs.
However, although the increases in $M_{\rm max}$ by rigid rotation of SSs are larger than that of NSs, at values of $M$ well below $M_{\rm TOV}$ the advantage of SSs over NSs regarding the increased of gravitational mass $M$ due to rotation over is not significant.

In fact, the gravitational energy releasing of an NS during spin-down is usually larger than that of an SS, since the shrinkage of the NS is larger.
Although this conclusion seems to favor NSs for providing gravitational energy to explain plateau emission, we still prefer SSs to NSs. The reason is that an NS as the remnant of binary neutron star mergers would not be long-lived. Although AP4 model for NSs could pass both the tests of the massive pulsars and tidal deformability during mergers~\citep{Annala2017}, $M_{\rm TOV}\sim 2.2 M_\odot$, which is well below the total mass of a known binary neutron star system inferred by~\citet{Antoniadis2016}.

\subsubsection{Spherical stretching due to rotation}
\label{subsubsec:stretch}

\begin{figure}[h!!!]
\centering
\includegraphics[width=9.0cm, angle=0]{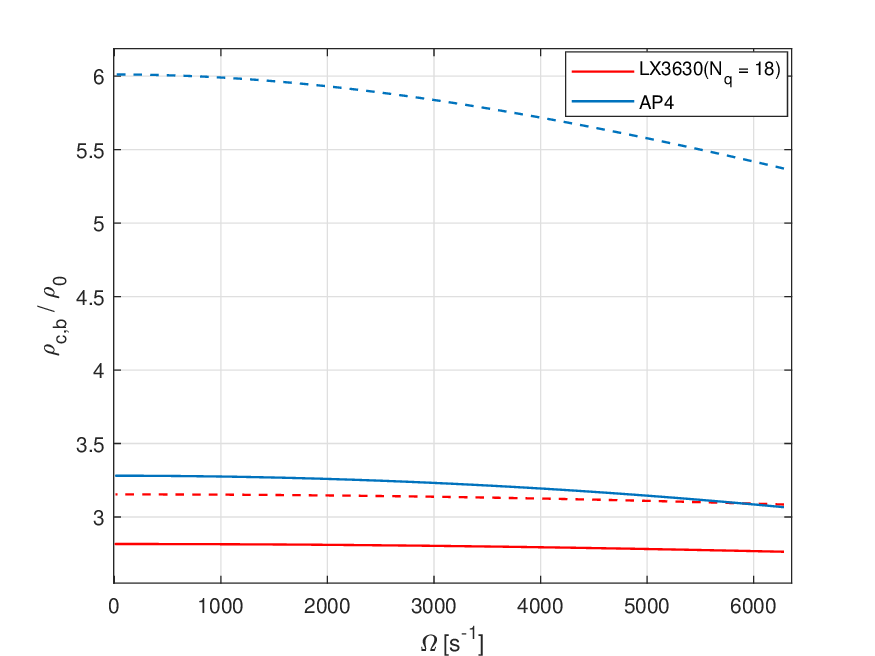}
\begin{minipage}[]{85mm}
\caption{The change of central baryon density $\rho_{\rm c,b}$ with $\Omega$ for $M_{\rm b}=1.6M_\odot$ (solid) and $M_{\rm b}=2.7M_\odot$ (dashed), where red lines represent results of SSs and blue lines represent results of NSs, respectively. $\rho_0$ is the saturated nuclear matter density.}
\end{minipage}
\label{fig_MR}
\end{figure}

From Fig.\ref{fig_rhocb} we can see that the between $\Omega=\Omega_{\rm c}$ and $\Omega=0$, the change in radius of an NS will be more than that of an SS.
The spherical stretching due to rotation has been discussed in~\citet{Gao2022}, which was found to be increasing towards the surface of the star for the case of NSs while be nearly unchanged through the star in the case of SSs.
To see the change of density due to rotation, we show explicitly here the change of central baryon density.
We plot the curves showing the change of central baryon density $\rho_{\rm c,b}$ with $\Omega$ under some given values of initial mass $M_0=M(\Omega=\Omega_0)$ in Fig.\ref{fig_MR}, with the initial angular frequency $\Omega_0=2\pi/(1\ \rm ms)$.
The red solid and dashed curves represent the results of SSs for $M_{\rm b}=1.6M_\odot$ (corresponding to the initial mass $M_0\simeq1.36M_{\odot}$) and $M_{\rm b}=2.7M_\odot$ (corresponding to $M_0\simeq2.16 M_{\odot}$), respectively.
For comparison, the results of NSs are shown by the blue solid and dashed curves for $M_{\rm b}=1.6M_\odot$ (corresponding to $M_0\simeq1.44M_{\odot}$) and $M_{\rm b}=2.7M_\odot$ (corresponding to $M_0\simeq2.22 M_{\odot}$), respectively.
%

From the change of $\rho_{\rm c,b}$ with $\Omega$, we can see that the spherical stretching of an NS is more significant than that of an SS with the same baryonic mass and the same initial spin frequency, especially when the initial mass $M_0$ is larger than $2M_\odot$.
Conversely, during spin-down, the shrinkage of an NS is more significant than that of an SS, especially in the case of binary merger remnants.
This may imply that a neutron stars with $M_0\sim 2.2 M_\odot$ as the remnant of binary NS merger, although being long-lived, would undergo phase transition during spin-down.
It would be interesting to explore the implication of such phase transition of massive or supramassive NSs, e.g. the energy released from the phase transition has been taken as the energy source related to GRBs' observations~\citep{Dai1998,Sarin2021PhDT}.
An SS, however, is close to the incompressible matter and would not undergo phase transition during spin-down.

It is worth noting that, if AP4 model applies to NSs both before and after mergers, the increase in central density during merger is much more significant than that during spin-down. The central density increases to almost twice as dense as that before merger, but only increase about 10 percent during spin-down. Therefore, if an NS with $M_0\sim2.2M_\odot$ as the remnant of the binary NS merger would undergo phase transition during spin-down, the phase transition would be more likely to occur during merger instead of during spin-down.

\section{Gravitational energy release of the binary merger remnants}
\label{sec:gravitation}

%
Because $M_{\rm TOV}$ values are high for strangeon stars, the remnants of the binary strangeon star mergers would probably not immediately collapse into black holes, and would be even long-lived.
Being spinning fast at the beginning, the remaining massive strangeon stars will undergo spin-down due to energy loss.
As shown in \S\ref{subsec:rotation}, the radius of a strangeon star will decrease as its angular frequency $\Omega$ decreases, so gravitational energy will be released during spin-down.
It will be shown that, although only a fraction of gravitational energy will be converted into the X-ray emissions, it may play an important role in the X-ray afterglow of SGRBs.

\subsection{Luminosity of gravitational energy}
\label{subsec:Lgrav}

The differential rotation may be a short-term process during the early stages of the merger remnants, and would be no longer important on the longer timescales for the afterglow.
In addition, we take the initial angular frequency $\Omega_0=2\pi/(1\ \rm ms)$, which satisfies the slow rotation condition $\Omega\ll \Omega_{\rm c} = \sqrt{GM/R^3}$.
Therefore, we can use the slow rotation approximation in the case of rigid rotation to derive the change of gravitational energy with time.

The gravitational energy of a relativistic star can be derived by
\begin{eqnarray}
E_{\rm grav}&=&Mc^2-M_{\rm P}c^2-E_{\rm kin}\nonumber\\&=&4\pi \int_0^R \rho (1-e^{\lambda})r^2{\rm d}r -\frac{1}{2}J\Omega
\end{eqnarray}
where the proper mass $M_{\rm P}$ is defined as $M_{\rm P}=4\pi \int_0^R \rho e^{\lambda}r^2{\rm d}r$, and the kinetic spin energy $E_{\rm kin}$ is related to the angular momentum $J$ by $E_{\rm kin}=J\Omega/2$.
The change of $E_{\rm grav}$ with $\Omega$ for a given $M_{\rm b}$ can be derived by the similar to the procedure used in \S\ref{subsec:rotation}, where the proper mass $M_{\rm P}$ is calculated to the second order of $\Omega$ by the procedure similar to that for deriving $M_{\rm b}$.
Using the slow rotation approximation to calculate $E_{\rm kin}$, the angular momentum $J$ is calculated to the first order of $\Omega$ by considering the rotational dragging of inertial frames, so $E_{\rm kin}$ is also calculated to the second order of $\Omega$.

The luminosity of gravitational energy can be derived by
\begin{equation}
L_{\rm grav}=\dot E_{\rm grav}=\frac{{\rm d}E_{\rm grav}}{{\rm d}\Omega}\frac{{\rm d} \Omega}{{\rm d} t}. \label{eq:L}
\end{equation}
Assuming that the spin-down is due to electromagnetic (EM) dipolar radiation and gravitational wave (GW) radiation, the change of $\Omega$ with time $t$ is~\citep{Shapiro1983}
\begin{equation}
\frac{{\rm d} \Omega}{{\rm d} t}=-\frac{B_p^2R^6\Omega^3}{6Ic^3}-\frac{32GI\epsilon^2\Omega^5}{5c^5}, \label{eq:spin-down}
\end{equation}
where $B_p$ is the dipolar field strength at the poles, $I$ is the moment of inertia, $R$ and $\epsilon$ are the radius and ellipticity of the star respectively.
Combining Eqs.(\ref{eq:L}) and (\ref{eq:spin-down}), we can get the luminosity of gravitational energy $L_{\rm grav}$.
It is worth noting that, in calculation of both $L_{\rm grav}$ and the luminosity of magnetic dipole radiation $L_{\rm em}=B_p^2R^6\Omega^4/(6c^3)$, the changes of $R$ and $I$ with $\Omega$ are taken into account.
$R$ is calculated by adding the spherical deformation to the second order of $\Omega$, as demonstrated in \S\ref{subsec:rotation}.
The corrections to $I=J/\Omega$ is also to the second order of $\Omega$, since the corrections to the angular moment $J$ for calculating $I$ is derived to the third order of $\Omega$~\citep{Hartle1973,Gao2022}.

\subsection{The role of $L_{\rm grav}$ in the X-ray afterglow of SGRBs}
\label{subsec:fit}

We investigate whether their gravitational energy release during spin-down could provide enough energy injection for the afterglow of SGRBs.
In a supernova explosion most of the gravitational energy will be taken away by neutrinos which are produced in phase transitions involving the weak interaction.
However, as shown in \S~\ref{subsubsec:stretch}, the shrinkage of an SS during the spin-down would not be large enough to cause a phase transition, so the loss of energy due to neutrinos would be unimportant.
Then how will the gravitational energy released due to the shrinkage of an SS be injected into the GRB firball?

Similar to the process of heating the solar corona and accelerating the solar wind by high-frequency Alfv\'{e}n waves~\citep{Tu1997,Kaghashvili1999}, the gravitational energy could be converted into kinematic one, and finally injected into the GRB fireball also by Alfv\'{e}n waves.
The oscillations and turbulence due to shrinkage of the star would lead to the magnetic reconnection and then generate Alfv\'{e}n waves to take away the kinetic energy (which comes from gravitational energy), like the oscillation-driven magnetospheric activity in pulsars~\citep{Lin2015}.
By this way, the form of converting gravitational energy into the fireball may also involve the processes similar to the kinetic-energy-dominated shell~\citep{Zhang2002} or a Poynting-flux-dominated outflow~\citep{Meszaros1997}.
Because the efficiency $\eta_{\rm g}$ for the gravitational energy to be converted into the X-ray emissions of GRB afterglow is unknown, all the complexities will be put into $\eta_{\rm g}$.

In the model which interprets the X-ray afterglow plateau emission in GRBs as being powered by the electromagnetic dipolar emission from millisecond magnetars, the efficiency $\eta_{\rm em}$ of converting the dipole spin-down luminosity to the observed luminosity should be considered.
However, $\eta_{\rm em}$ is derived by fitting observational data because its prior value is hard to be calculated, and the fitting results are usually different.
Although some simulations suggest that $\eta_{\rm em}=0.4-1$~\citep{Gao2016}, a more detailed study indicates that the X-ray radiation efficiency depends strongly on the saturation Lorentz factor and the typical value is of order $10^{-2}$~\citep{Xiao2019}.
Similarly, the efficiency $\eta_{\rm g}$ of converting the gravitational energy luminosity to the observed X-ray luminosity should also be considered, but how to determine $\eta_{\rm g}$ is still a problem.
The data points of each SGRB we choose are not enough to give good fitting with more than three free parameters, we can only choose either $\eta_{\rm g}$ or $\eta_{\rm em}$ as a free parameter. We find that the fitting results of setting $\eta_{\rm g}=\eta_{\rm em}$ will not differ significantly from that of setting $\eta_{\rm em}=0$, so to highlight the role of gravitational energy we set $\eta_{\rm em}=0$, except two cases in which we set $\eta_{\rm g}=\eta_{\rm em}$ to avoid $\eta_{\rm g}>1$, as will be shown latter.
In addition, because the spin-down due to GW radiation would not be important in the afterglow phase~\citep{Zhang2001}, we neglect the second term of Eq.(\ref{eq:spin-down}).

To test the validity of our scenario, we choose some SGRBs from~\citet{Stratta2018} which fits a sample of GRB X-ray afterglows by assuming that the plateau emission is powered by the spin-down luminosity of millisecond magnetars.
They derived values of $\chi^2/\rm d.o.f$ of the fitting results which can be used to be compared with our results.
Because we do not use model or quantitative selection criteria to identify the plateau phase, we choose the ones which have obvious plateaus, and identify the flat part of the data to be the plateau phase to begin our fitting.
Among the ten SGRBs fitted in~\citet{Stratta2018} we choose six ones, 051221A, 060614, 061201, 070714B, 070809 and 090510, which have red-shifts and obvious plateaus. Beside them, we choose another two SGRBs from the Swift GRB sample obvious plateaus, 130603B and 140903A.

For a given burst in {\it Swift} data~\citep{Evans2007,Evans2009}, the source rest frame luminosity is derived from the flux $F(t)$ at the time $t$ by $L(t)=4\pi D_L(z)^2F(t)$,
where $D_L(z)$ is the luminosity distance at redshift $z$. For the correction from $1-10^4$ keV in the burst rest frame to the observed band, the X-ray luminosity derived in our model is divided by the factor $k_c$.
The redshifts of the eight SGRBs are from Table 1 of~\citet{Kisaka2017} (and references therein).
We adopt the values of cosmological parameters used in~\citet{Komatsu2009} to get $D_L(z)$, and $k_c$  is derived by the method used in~\citet{Bloom2001}.

In the MCMC fits of different SGRBs, we assume $M_{\rm b}\simeq 3.1 M_\odot$, which corresponds to $M_0\simeq 2.36 M_\odot$ under the fitted values of the initial spin period $P_0$.
Certainly, the mass range for remnants in binary star mergers is unknown.
The total mass of the binary system associated with GW170817 which is probably larger than $2.7M_\odot$~\citep{GW170817} and the masses of the known binary neutron star systems~\citep{Antoniadis2016} can infer that, the initial mass $M_0\simeq 2.36 M_\odot$ for remnants in binary star mergers could be reasonable.
We find that the change of $M_0$ from $2.2M_\odot$ to $M_0=2.5M_\odot$ would not make significant effects.

\begin{figure}[h!!!]
\centering
\includegraphics[width=6.5cm, angle=0]{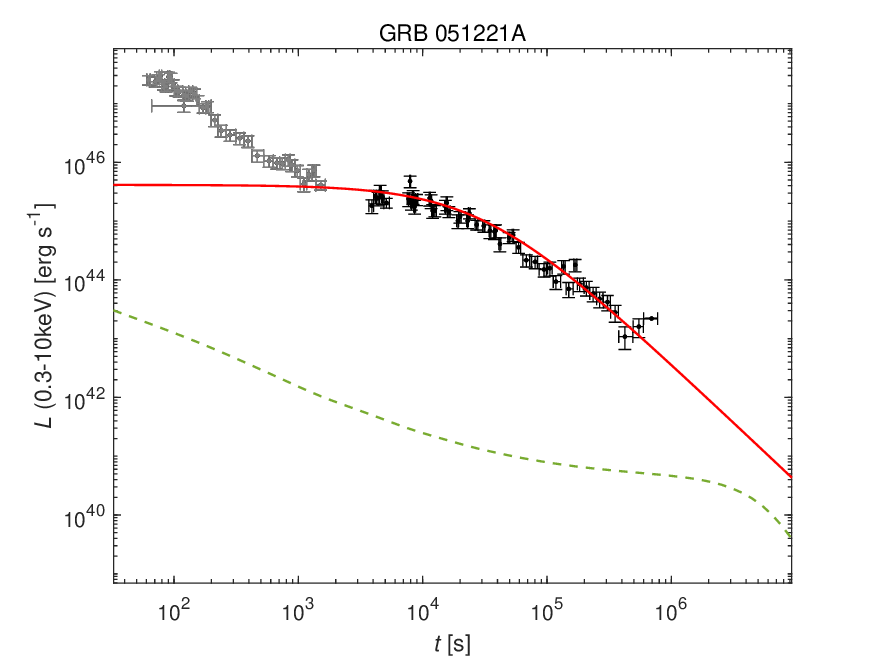}
\includegraphics[width=6.5cm, angle=0]{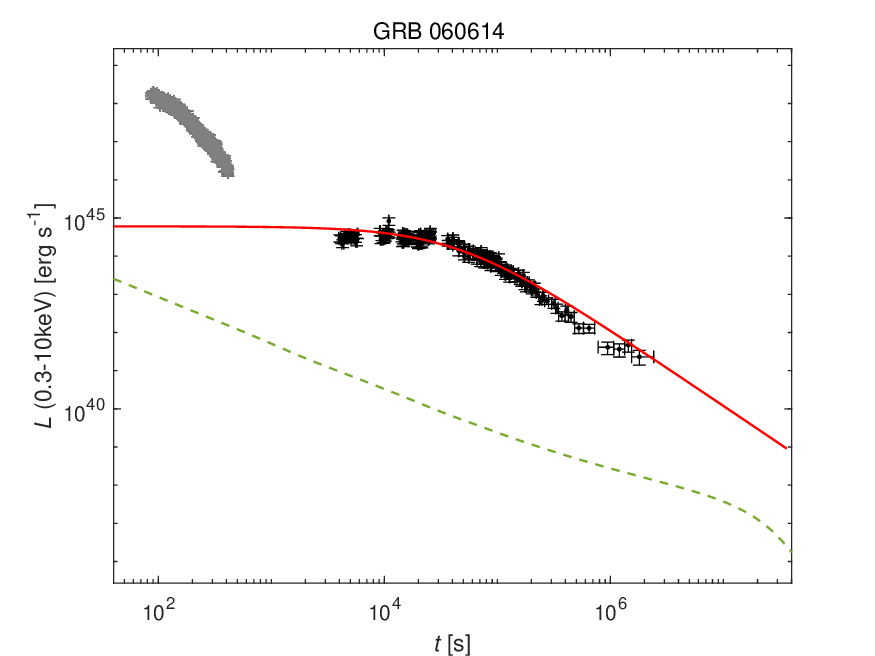}
\includegraphics[width=6.5cm, angle=0]{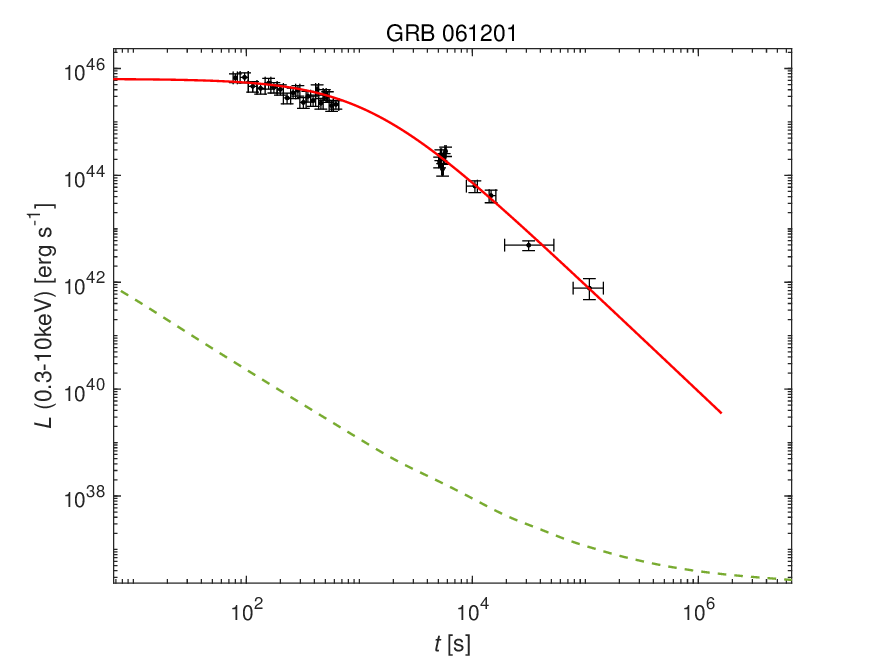}
\includegraphics[width=6.5cm, angle=0]{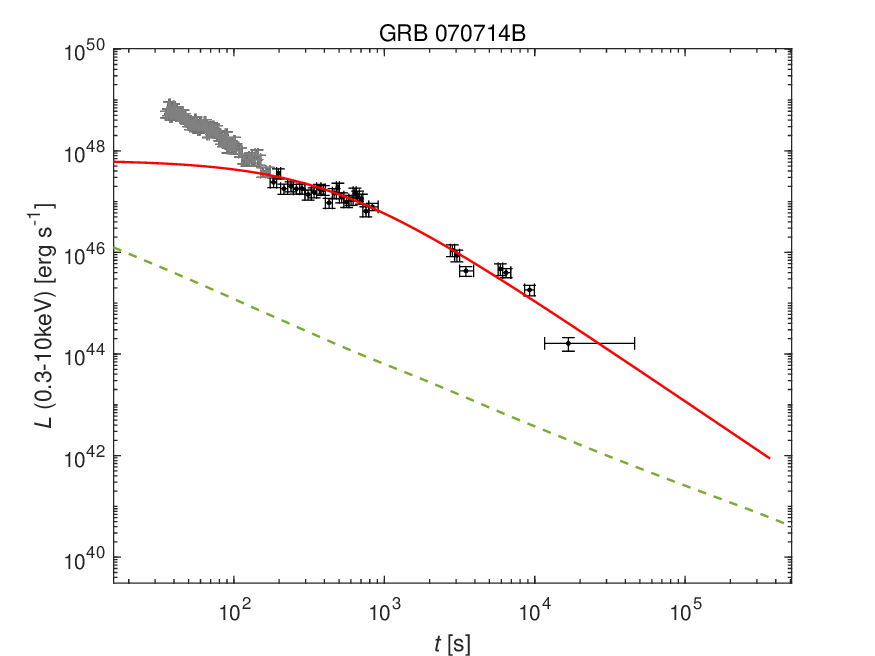}
\includegraphics[width=6.5cm, angle=0]{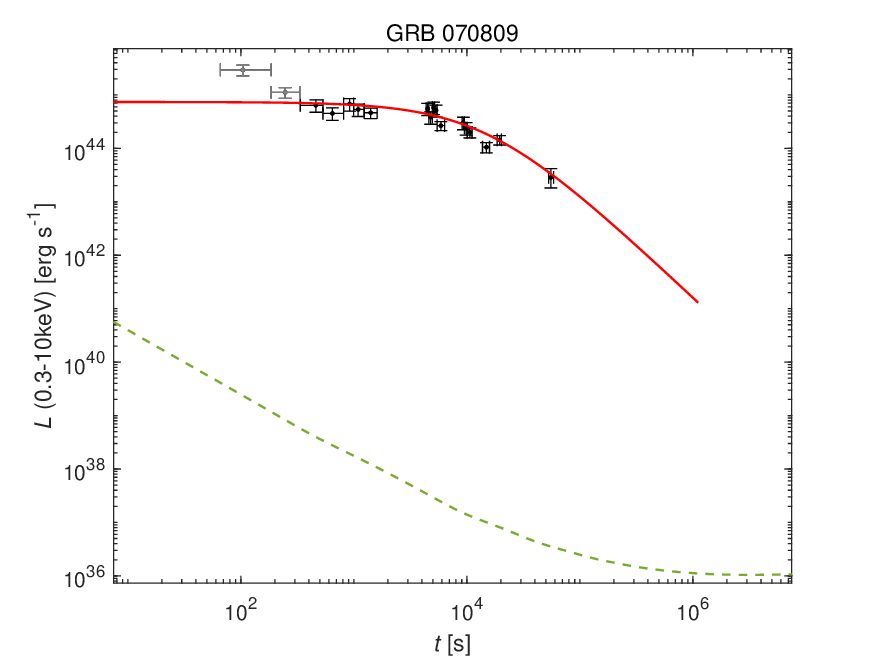}
\includegraphics[width=6.5cm, angle=0]{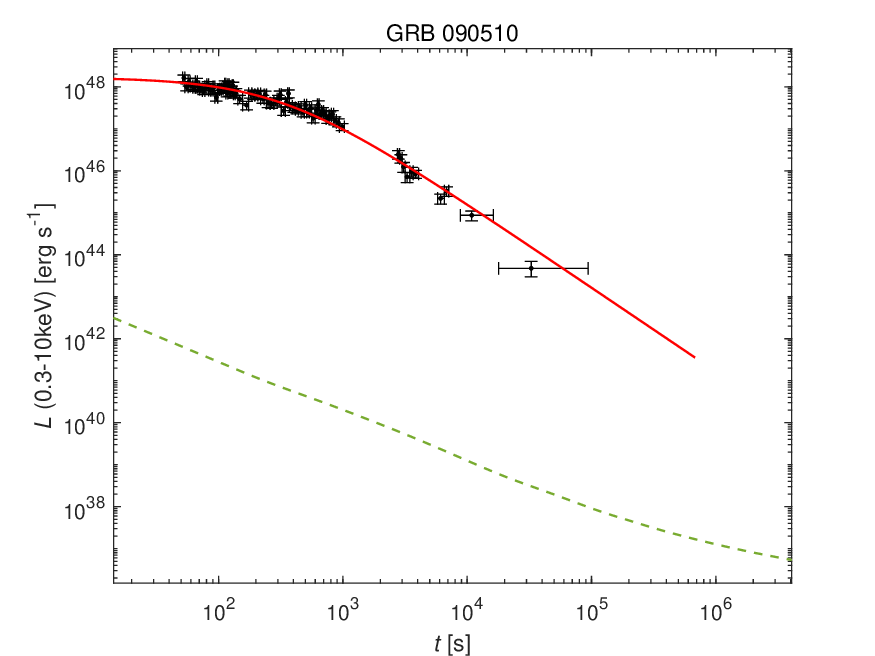}
\includegraphics[width=6.5cm, angle=0]{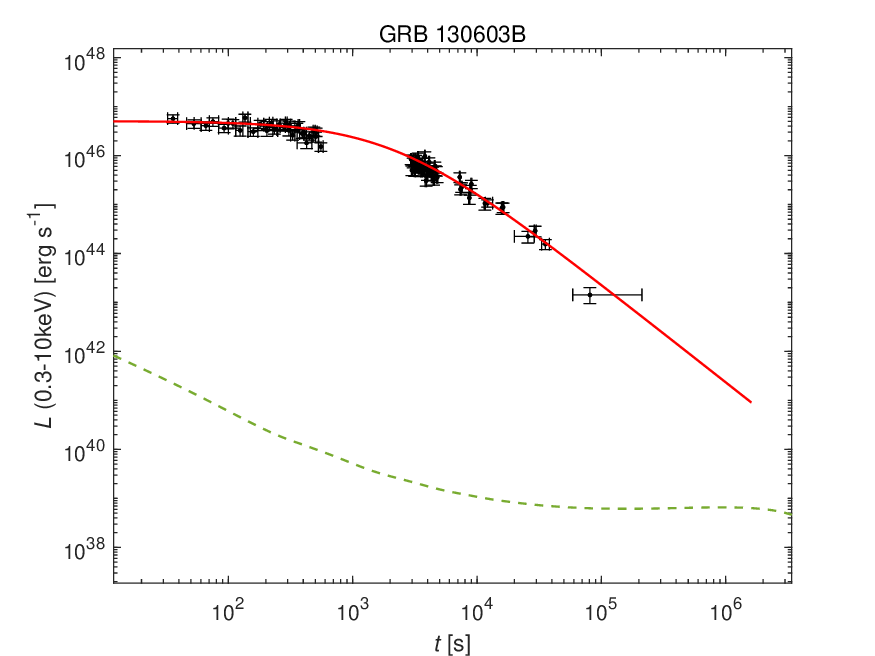}
\includegraphics[width=6.5cm, angle=0]{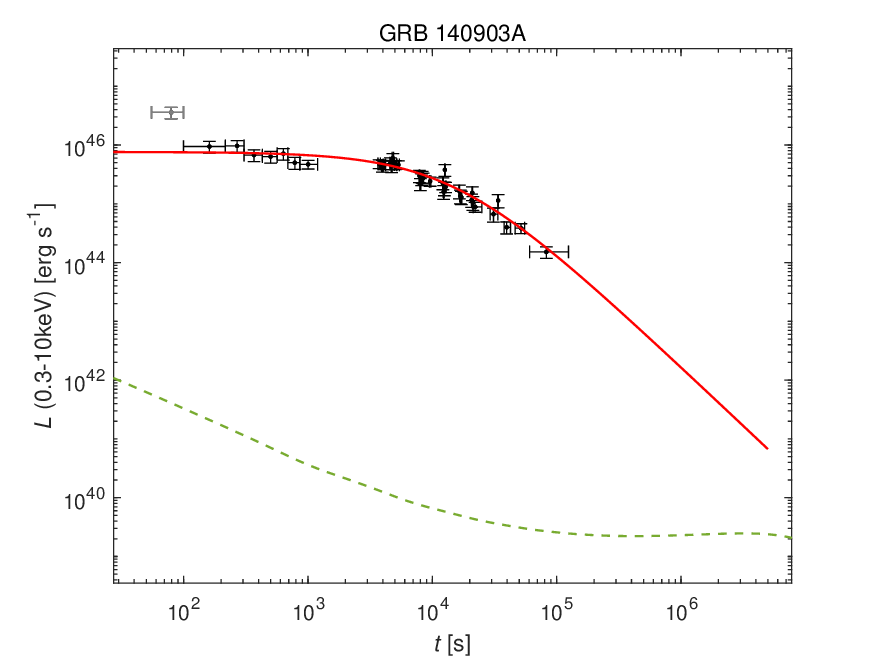}
\caption{X-ray afterglow light curves of eight SGRBs with our best-fit results, fitted to the black points which begin from the flat part of the data. The afterglow component from the interaction between the jet and interstellar medium are shown by dashed lines.}
\label{fig:fit}
\end{figure}

The afterglow component from the interaction between the jet and interstellar medium should also be considered. We use the Python package \verb"afterglowpy"~\citep{Ryan2020} which utilizes semianalytic approximations to the jet evolution and synchrotron emission to calculate afterglow light curves with structured jets, taking into account relativistic beaming effect. In our calculations, the Gaussian jet model is used to calculate the contribution of interaction between the jet and interstellar medium, assuming that fractions of post-shock energy in radiating electrons $\epsilon_e=0.04$ and magnetic fields $\epsilon_B=10^{-4}$~\citep{Ryan2020}, the jet half-opening angle $\theta_c=6.87^{\circ}$~\citep{Fong2015}, and the truncation angle $\theta_w=5\theta_c$. The values of the number density of interstellar medium $n_0$ and the electron power-law distribution index $p$ are chosen from~\citet{Fong2015} for each SGRB. The isotropic equivalent energy of the blast wave $E_0=10^{52.2}$ ergs~\citep{CaoXF2023}, and the viewing angle $\theta_v$ can be derived from the photon index and peak flux for the Gaussian jet. After obtaining the flux at 3 keV under the above parameters using \verb"afterglowpy", the flux at 0.3-10 keV can be derived by the method of~\citet{Gehrels2008}. The results are shown by dashed lines in Fig.\ref{fig:fit}, which indicate that compared with the internal energy injection, the afterglow component from the interaction between the jet and interstellar medium would not be important.

\begin{table}[t]
    \centering
    \caption{Best-fit results of ours (left) and~\citet{Stratta2018} (right)}
    \label{tab:fit}
    \begin{tabular}{ccccc||cc} 
        \hline
        SGRB & $B_p\ (10^{15}\ {\rm G})$ & $P_0\ (\rm ms)$ & $\eta_{\rm g}$ & $\chi^2/{\rm d.o.f.}$ & $B_p\ (10^{15}\ \rm G)$ & $\chi^2/{\rm d.o.f.}$\\[0.2cm]
        \hline
        051221A & $1.69^{+0.19}_{-0.34}$ & $5.29^{+0.63}_{-1.06}$ & $0.69^{+0.18}_{-0.25}$ &  1.1017 & $10.4\pm 0.9$ & 1.3213\\[0.2cm]
        060614 & $0.88^{+0.13}_{-0.13}$ & $3.33^{+0.51}_{-0.50}$ & $0.11^{+0.04}_{-0.03}$ &  1.9674 & $15.6\pm 0.5$ & 1.0821\\[0.2cm]
        061201 & $4.17^{+0.54}_{-0.90}$ & $2.60^{+0.34}_{-0.54}$ & $0.14^{+0.04}_{-0.05}$ &  1.2937 & $20.6\pm 1.5$ & 1.2417\\[0.2cm]
        070714B & $8.73^{+1.31}_{-2.04}$ & $3.23^{+0.50}_{-0.75}$ & $0.58^{+0.19}_{-0.24}$ &  2.1809 & $36.9\pm 4.0$ & 1.9500\\[0.2cm]
        070809 & $4.98^{+0.80}_{-1.20}$ & $10.94^{+1.53}_{-2.58}$ & $0.63^{+0.21}_{-0.26}$ & 1.1828  & $5.8\pm 1.1$ & 1.2615\\[0.2cm]
        090510 & $6.40^{+0.86}_{-1.52}$ & $2.05^{+0.28}_{-0.48}$ & $0.25^{+0.07}_{-0.11}$ & 1.9234 & $11.6\pm 0.5$ & 1.1914\\
        \hline
    \end{tabular}
\end{table}

\begin{figure}[h!!!]
\centering
\includegraphics[width=5cm, angle=0]{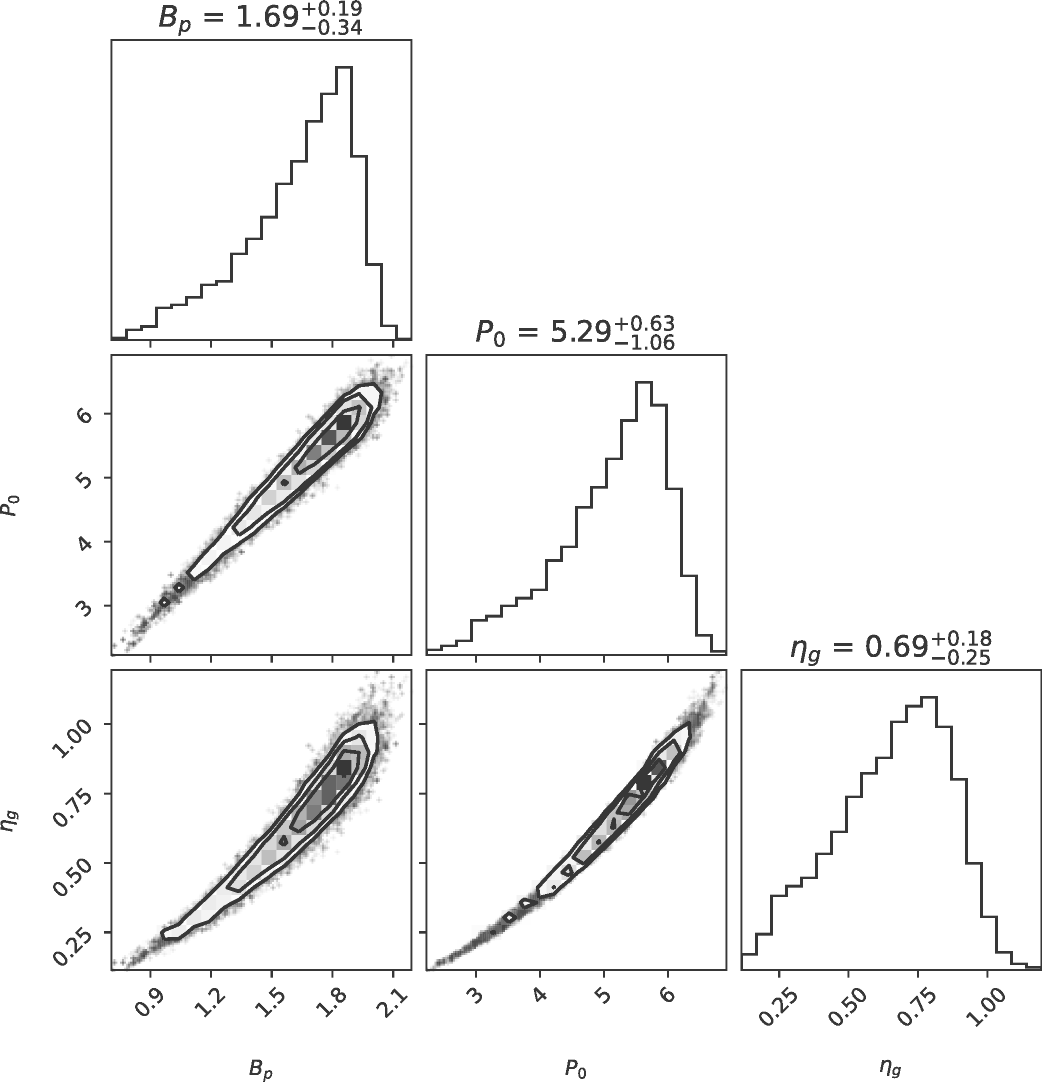}
\includegraphics[width=5cm, angle=0]{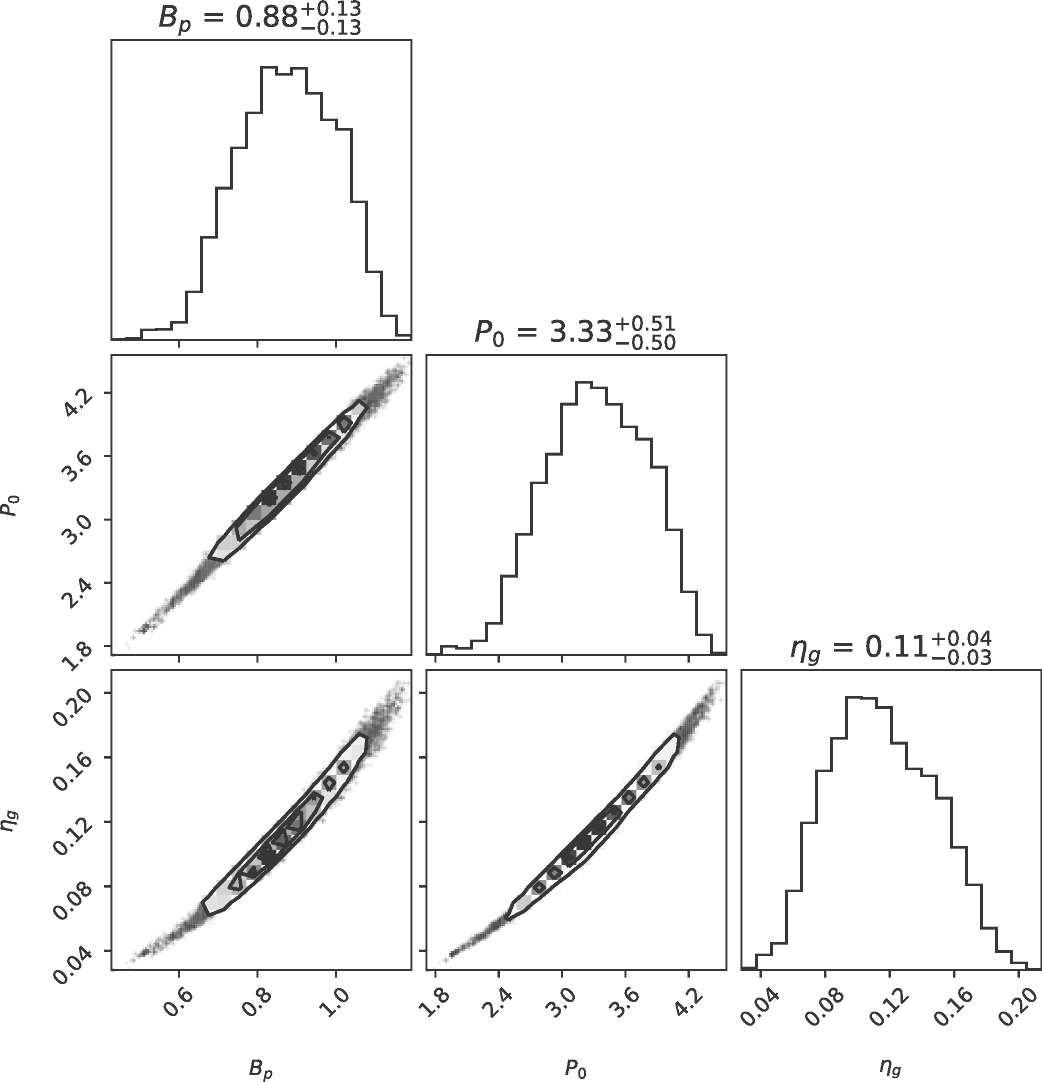}
\includegraphics[width=5cm, angle=0]{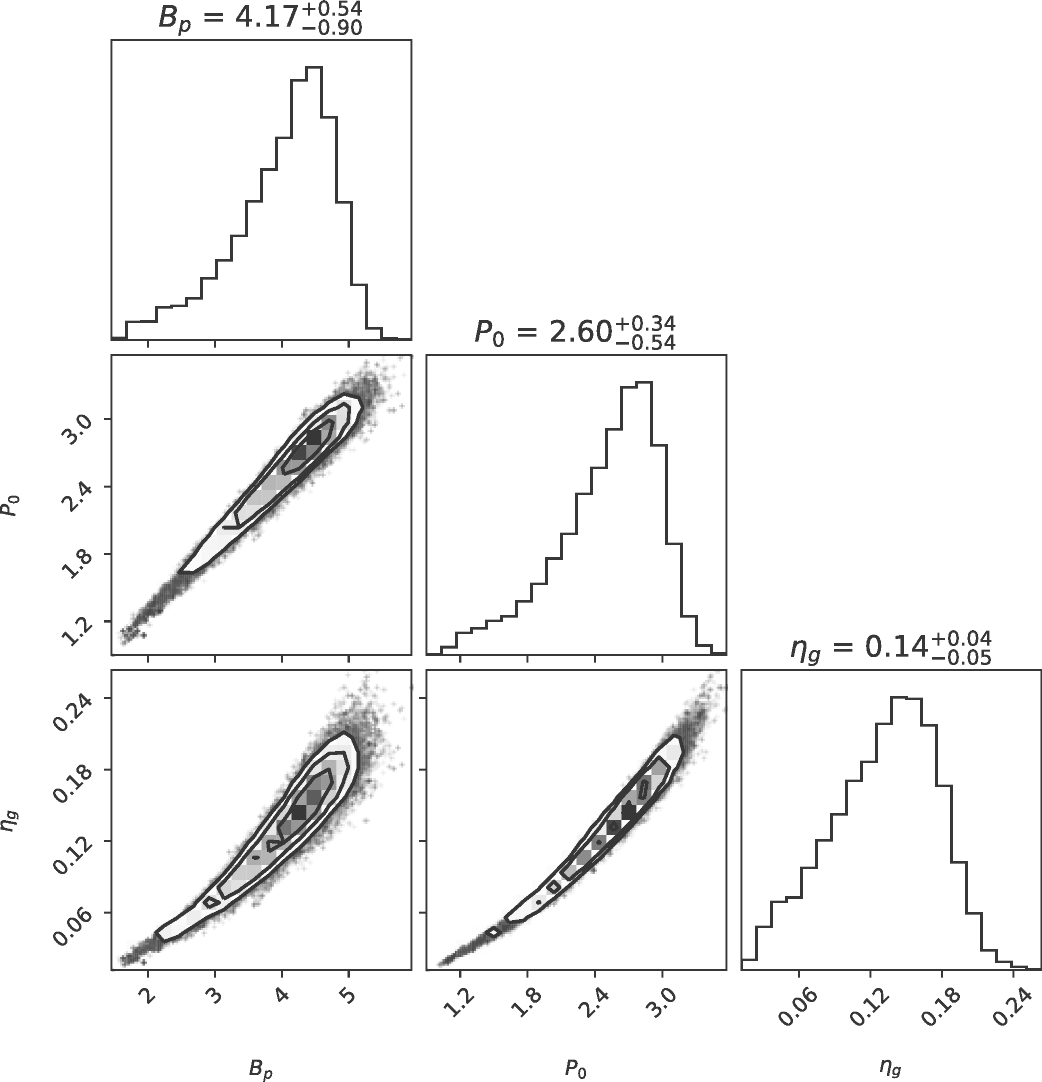}
\includegraphics[width=5cm, angle=0]{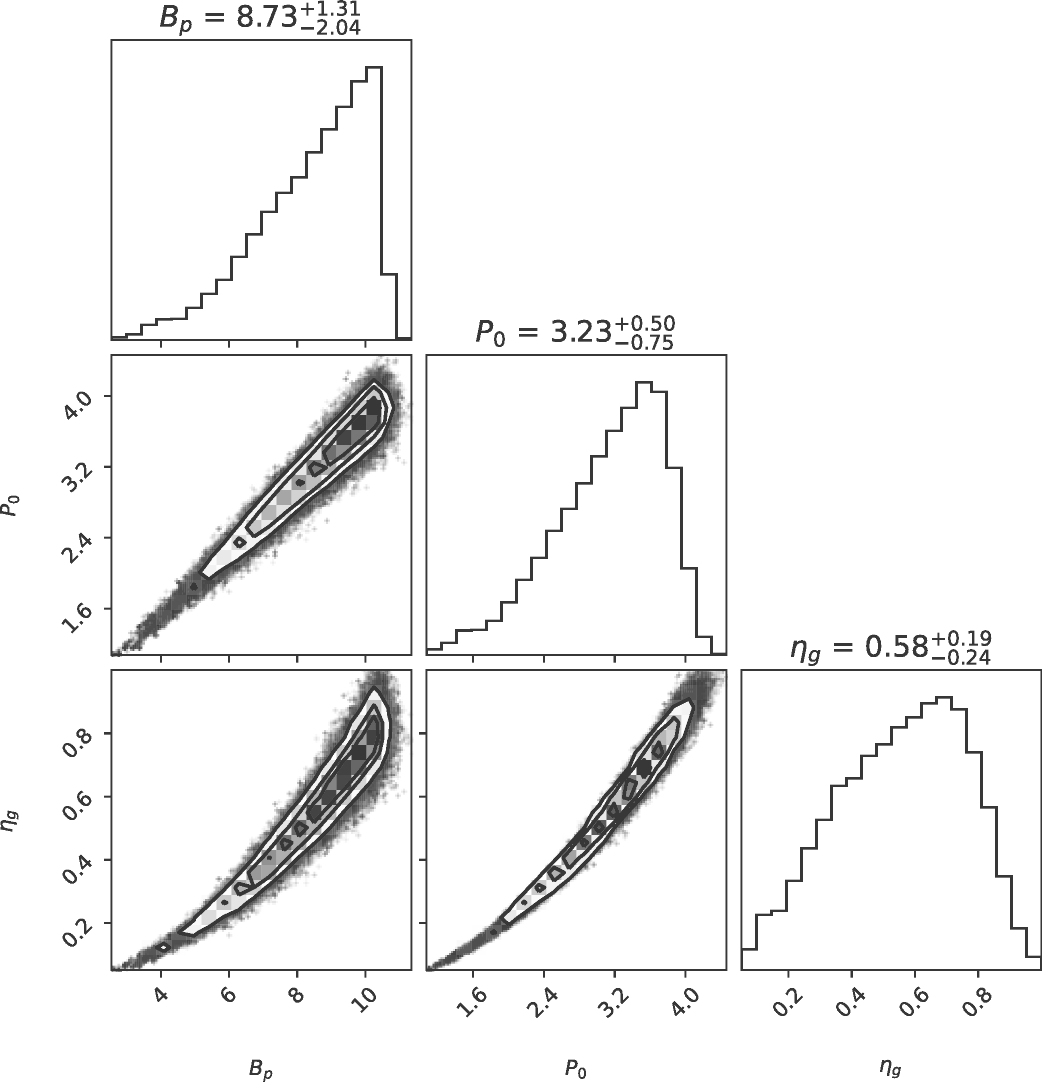}
\includegraphics[width=5cm, angle=0]{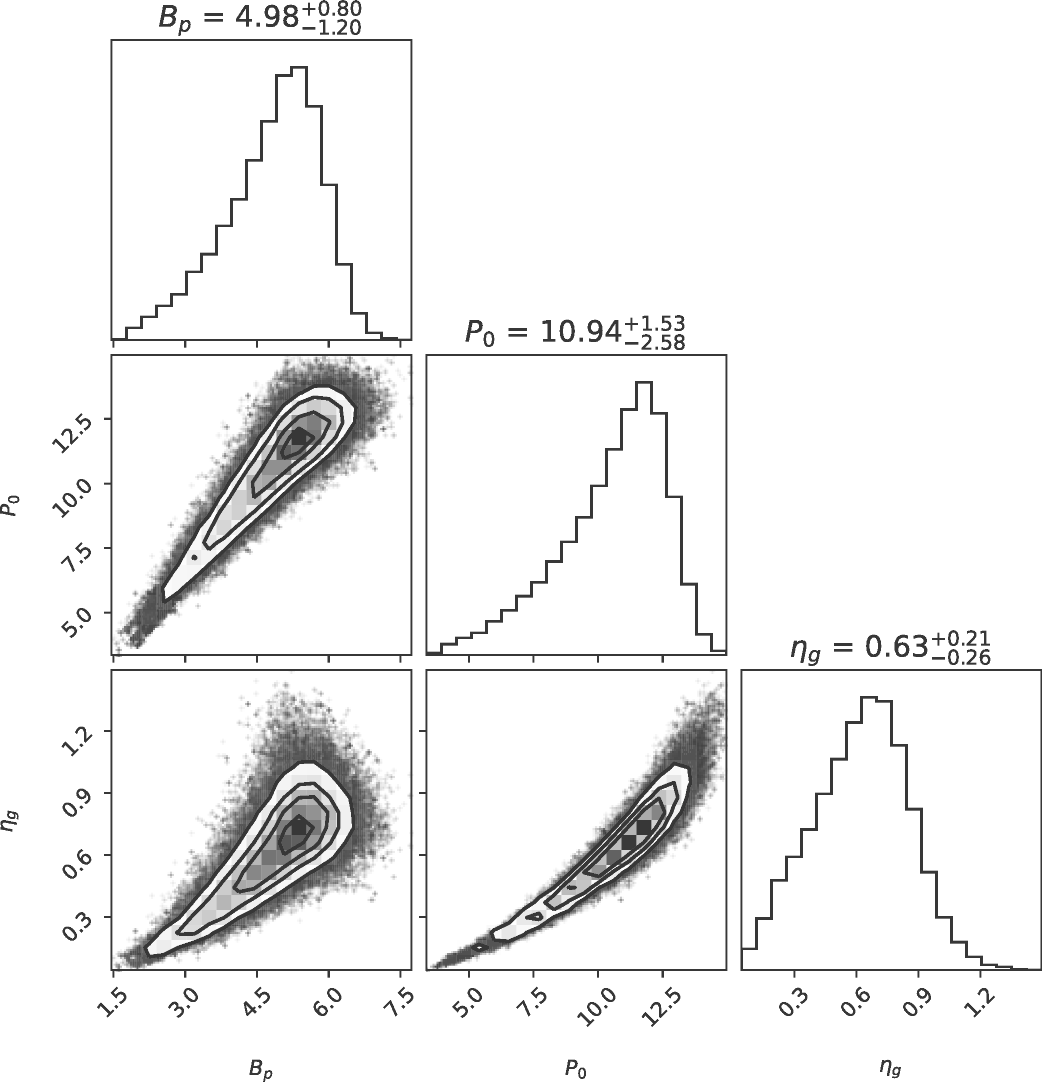}
\includegraphics[width=5cm, angle=0]{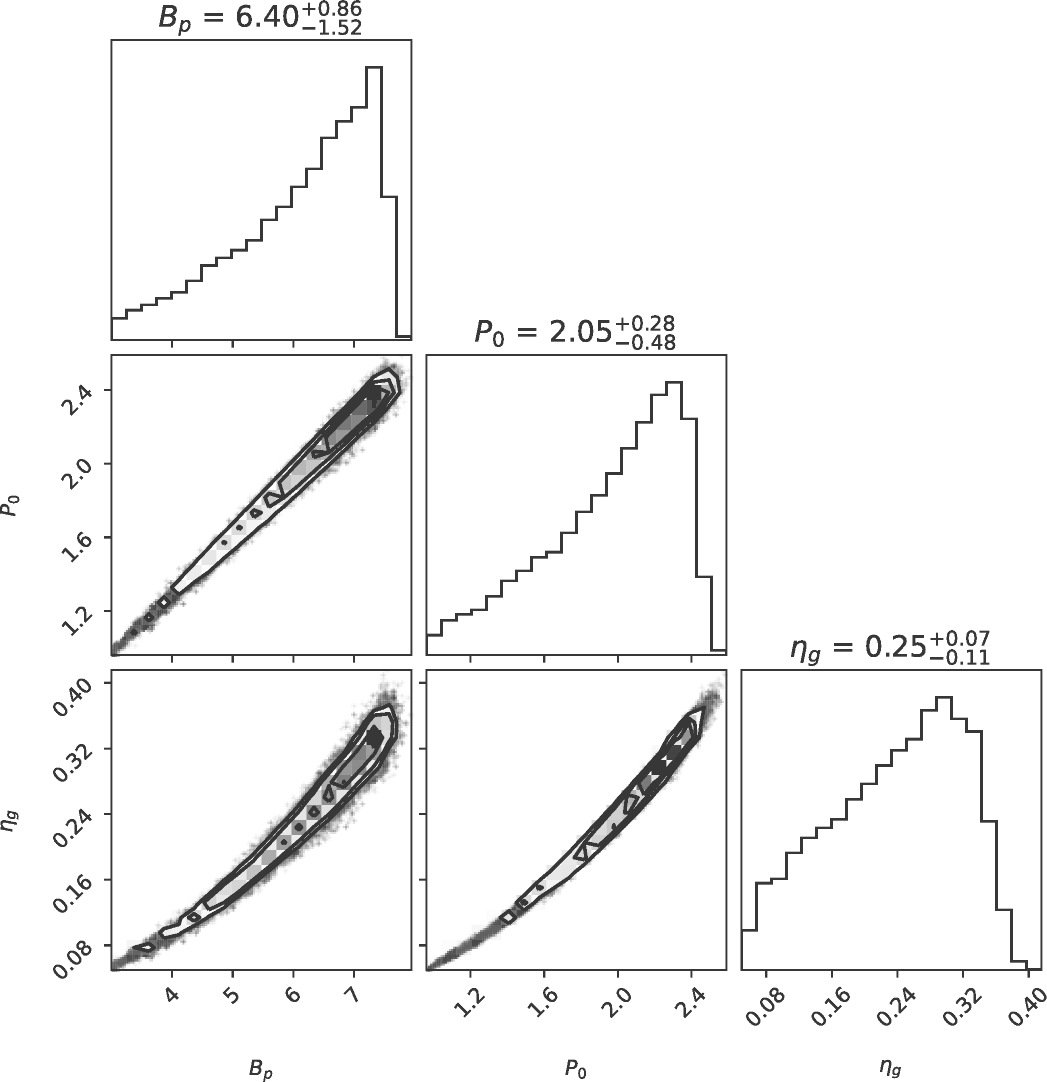}
\includegraphics[width=5cm, angle=0]{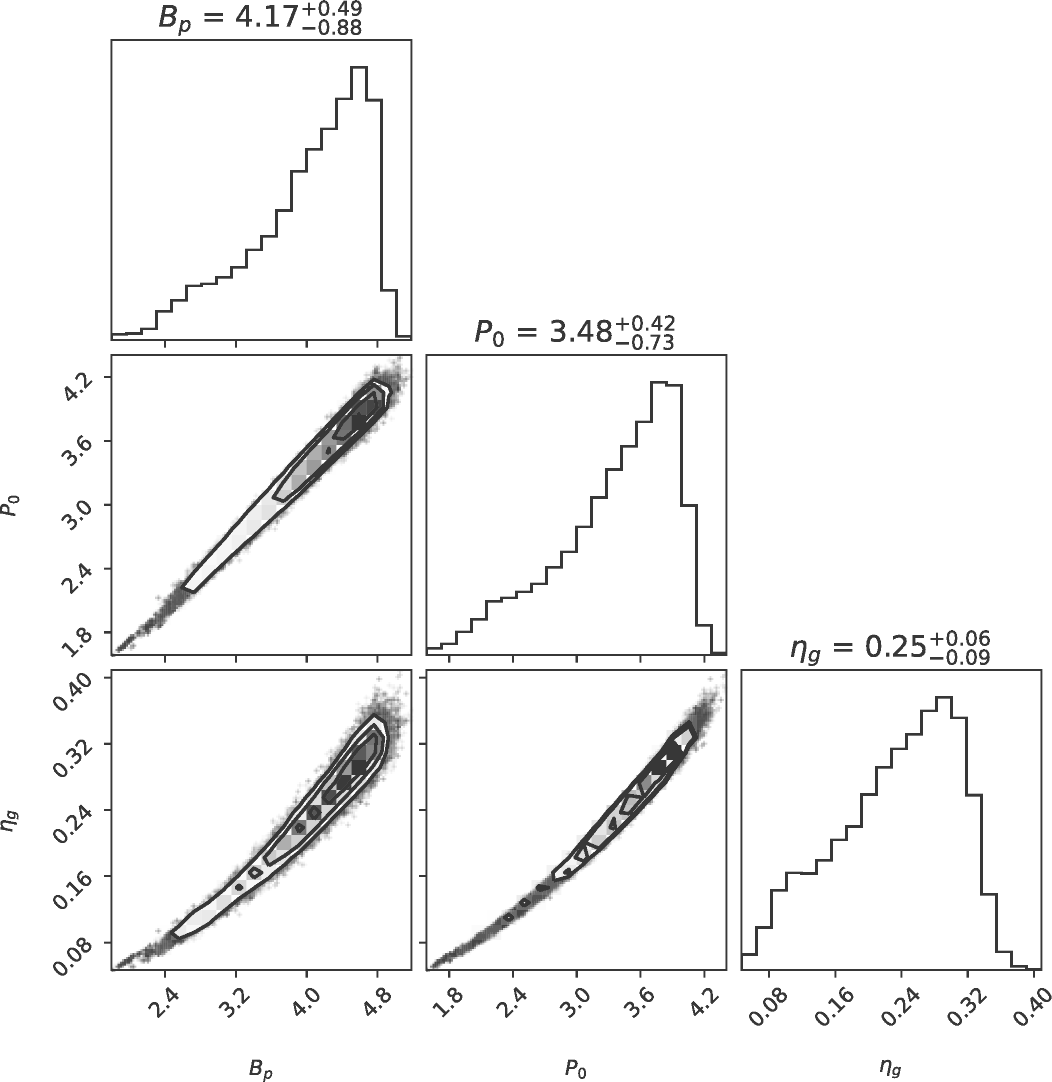}
\includegraphics[width=5cm, angle=0]{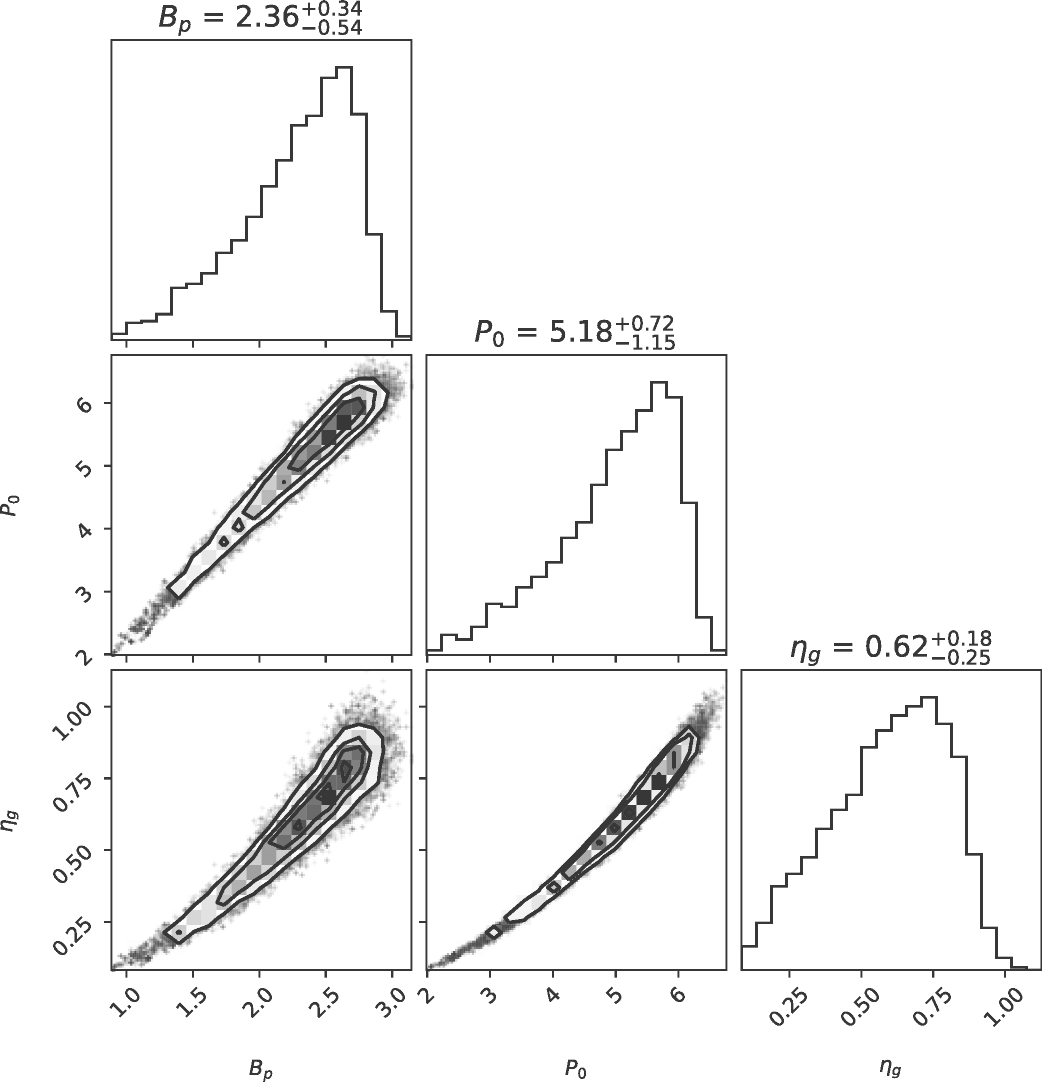}
\caption{The histograms and contours for MCMC fits of the X-ray afterglow light curves of eight SGRBs in Fig.\ref{fig:fit}.}
\label{fig:corner}
\end{figure}

We use the MCMC (Markov Chain Monte Carlo) method to fit three parameters including $B_p$, $P_0$ and $\eta_{\rm g}$.
The comparison of the fitting results of ours  and that of~\citet{Stratta2018}  for six SGRBs  are shown in Table~\ref{tab:fit}.
The X-ray light curves ($1-10^4$ keV) of eight SGRBs and the histograms and contours for MCMC fits are shown in Fig.\ref{fig:fit} and Fig.\ref{fig:corner}, respectively.
Because we find that the fitting results of setting $\eta_{\rm g}=\eta_{\rm em}$ will not differ significantly from that of setting $\eta_{\rm em}=0$, we choose to fix $\eta_{\rm em}$ to avoid introducing one more parameter. To highlight the role of gravitational energy we set $\eta_{\rm em}=0$ for 051221A, 060614, 061201, 070714B, and 070809. For 090510 and 140903A, we set $\eta_{\rm g}=\eta_{\rm em}$ to avoid $\eta_{\rm g}>1$.
From the fitting results we can see that, for the six SGRBs from~\citet{Stratta2018}, the $\chi^2/\rm d.o.f$ values of our scenario not much larger than that of~\citet{Stratta2018}, indicating that our scenario is at least not bad.
In addition, the values of $B_p$ are much smaller than that of expected in~\citet{Stratta2018}.

\section{Conclusions and discussions}
\label{sec:conclusion}

The effects of rotation on the structure of pulsar-like compact stars would have interesting consequences and provide tests for the EoS models.
Under the hypothesis that pulsar-like compacts stars are SSs, whose EoS is described by the Lennard-Jones model, we study rigidly rotating SSs in the slow rotation approximation.
Although we only choose the EoS of LX3630 and $N_{\rm q}=18$ to perform our calculations, the other forms of EoS under different parameters may not bring qualitative differences.
We find that rotation can significantly increase the maximum mass of a stable SS.
The SS with mass $M$ larger than $M_{\rm TOV}$ by approximately 9\% can still be stable or long-lived during spin-down with unchanged baryonic mass $M_{\rm b}$.
Considering that $M_{\rm TOV}$ of SSs can be much larger than $2.5M_{\odot}$ in a wide range of parameter space, it is very likely that the remnants of binary SS mergers are long-lived massive SSs.

To explore the consequences of rotating massive SSs, we investigate whether their gravitational energy release during spin-down could provide enough energy injection for the afterglow of SGRBs.
We derive the luminosity of gravitational energy releasing, where the spin-down is due to magnetic dipolar radiation and the changes of radius $R$ and the moment of inertial $I$ with angular frequency $\Omega$ are taken into account.
The X-ray light curves can be derived by assuming that a fraction of gravitational energy releasing contributes to X-ray luminosity.
By fitting X-ray afterglow of six SGRBs in~\citet{Stratta2018} who have red-shifts and obvious plateaus, we find that the gravitational energy released by long-lived massive strangeon stars could provide an alternative energy source for the plateau emission of X-ray afterglow.
Our fitting results show that the magnetic dipole field strength of the remnants can be much smaller than that of expected in the magnetar scenario.
The fitting results of our scenario seem not bad compared to the magnetar scenario which is much more sophisticated than ours.

The followings are some discussions.
Although the plateau of the X-ray afterglow of SGRBs is widely accepted as being powered by the electromagnetic dipolar emission from millisecond magnetars, we demonstrate in this paper that gravitational energy could provide an alternative energy source.
To avoid the complexity in the details of the millisecond magnetar origin scenario, we choose to fix $\eta_{\rm em}$ to be either 0 or equal to $\eta_{\rm em}$.
It is expected to find a reasonable way to combine both of them to account for the X-ray afterglow emission of SGRBs, by fitting a larger sample of SGRBs.

Certainly, we use some simplifications and assumptions to get our results.
The star is assumed to be rotating rigidly, and the slow rotation approximation is used to calculate its structure.
Moreover, how the gravitational energy can be injected into GRB fireball to power the afterglow emission is unknown.
Here we assume that a fraction of gravitational energy could be converted into kinematic one and finally injected into the GRB fireball by Alfv\'{e}n waves, which could be generated by oscillations and turbulence due to shrinkage of the star, and we put all the ignorance into the efficiency $\eta_{\rm g}$.
Similar to the efficiency $\eta_{\rm em}$ in the magnetar scenario, $\eta_{\rm g}$ would not be a constant and would depend on many factors, such as the injected luminosity and the process of injection.
The improve version of our scenario in the future by fitting a larger sample of SGRBs would be promising.

Some further investigations about GRB afterglow are expected.
We only consider in this paper the afterglow of the SGRBs instead of the long GRBs (LGRBs), which are generally believed to originate from the supernova explosions whose remnants are pulsar-like compact stars or black holes.
The millisecond magnetars, if the magnetic dipole field could be high enough, are generally accepted to be the engine of X-ray afterglow plateaus for both LGRBs and SGRBs.
However, for the case of LGRBs whose remnants are stable compact stars with mass about $1.4M_\odot$, the gravitational energy released during spin-down alone may not be large enough to account for the X-ray afterglow plateaus, no matter the remnants are SSs or NSs.
The real process of energy injection into the GRB afterglow would be complex, and the gravitational energy would only be a part of the available energy sources.

In addition, we only consider the long-lived remnants of binary mergers and their implications on the X-ray afterglow plateaus of SGRBs.
Some X-ray afterglows of SGRBs show rapid decay after the plateau phase, in which case the plateau is interpreted to be powered by the supramassive remnants and the rapid decay is thought to be the signal of collapse into BHs.
The study about the distribution of the break time (i.e. collapse time) of X-ray plateau in SGRBs could provide test for EoS models of NSs and QSs~\citep{LiA2016}.
For SS model, the implication for the break time of X-ray plateaus in SGRBs is worth exploring.
On one hand, to calculate the break time we should know the mass-distribution of the binary neutron stars via population synthesis and more detailed observations on SGRBs, which at present have many uncertainties.
And on the other hand, the sudden decrease of X-ray light curves may not imply the collapse into black holes, since the decay of $L_{\rm grav}$ could be rapid even for a long-live massive SS.
As the first attempt to exploring the related issues, in this paper we only consider the case of long-lived remnants and try to explain the observed luminosity of X-ray plateau.
Further works about how to test different EoS models by observations of GRBs are expected.

\normalem
\begin{acknowledgements}
We would like to thank Dr. Yong Gao and Prof. Yun-Wei Yu for useful discussions, and to thank the anonymous referee for comments.
This work makes use of data supplied by the UK Swift Science Data Centre at the University of Leicester.
This work is supported by National SKA Program of China (No. 2020SKA0120300),
the Outstanding Young and Middle-aged Science and Technology Innovation Teams from Hubei colleges and universities (No. T2021026),
the Young Top-notch Talent Cultivation Program of Hubei Province,
and the Key Laboratory Opening Fund (MOE) of China (grant No. QLPL2021P01).
\end{acknowledgements}


\begin{thebibliography}{67}
\providecommand\natexlab[1]{#1}
\providecommand\JournalTitle[1]{#1}

\bibitem[Abbott {et~al.}(2017)]{GW170817}
Abbott, B.~P., {et~al.} 2017, Phys. Rev. Lett., 119, 161101

\bibitem[{Akmal} \& {Pandharipande}(1997)]{AP1997}
{Akmal}, A., \& {Pandharipande}, V.~R. 1997, \prc, 56, 2261

\bibitem[Annala {et~al.}(2018)]{Annala2017}
Annala, E., Gorda, T., Kurkela, A., \& Vuorinen, A. 2018, Phys. Rev. Lett.,
  120, 172703

\bibitem[{Antoniadis} {et~al.}(2016)]{Antoniadis2016}
{Antoniadis}, J., {Tauris}, T.~M., {Ozel}, F., {et~al.} 2016, arXiv e-prints,
  arXiv:1605.01665

\bibitem[{Benhar} {et~al.}(2005)]{Benhar2005}
{Benhar}, O., {Ferrari}, V., {Gualtieri}, L., \& {Marassi}, S. 2005, \prd, 72,
  044028

\bibitem[{Berti} \& {Stergioulas}(2004)]{Berti2004}
{Berti}, E., \& {Stergioulas}, N. 2004, \mnras, 350, 1416

\bibitem[{Berti} {et~al.}(2005)]{Berti2005}
{Berti}, E., {White}, F., {Maniopoulou}, A., \& {Bruni}, M. 2005, \mnras, 358,
  923

\bibitem[{Bloom} {et~al.}(2001)]{Bloom2001}
{Bloom}, J.~S., {Frail}, D.~A., \& {Sari}, R. 2001, \aj, 121, 2879

\bibitem[{Cao} {et~al.}(2023)]{CaoXF2023}
{Cao}, X.-F., {Tan}, W.-W., {Yu}, Y.-W., \& {Zhang}, Z.-D. 2023, arXiv
  e-prints, arXiv:2306.16795

\bibitem[{Dai} {et~al.}(2011)]{DLX2011}
{Dai}, S., {Li}, L., \& {Xu}, R. 2011, Science China Physics, Mechanics, and
  Astronomy, 54, 1541

\bibitem[{Dai} \& {Lu}(1998)]{Dai1998}
{Dai}, Z.~G., \& {Lu}, T. 1998, \aap, 333, L87

\bibitem[{Dai} {et~al.}(2006)]{DaiZG2006}
{Dai}, Z.~G., {Wang}, X.~Y., {Wu}, X.~F., \& {Zhang}, B. 2006, Science, 311,
  1127

\bibitem[{Eichler} {et~al.}(1989)]{Eichler1989}
{Eichler}, D., {Livio}, M., {Piran}, T., \& {Schramm}, D.~N. 1989, \nat, 340,
  126

\bibitem[{Evans} {et~al.}(2007)]{Evans2007}
{Evans}, P.~A., {Beardmore}, A.~P., {Page}, K.~L., {et~al.} 2007, \aap, 469,
  379

\bibitem[{Evans} {et~al.}(2009)]{Evans2009}
{Evans}, P.~A., {Beardmore}, A.~P., {Page}, K.~L., {et~al.} 2009, \mnras, 397,
  1177

\bibitem[{Fong} {et~al.}(2015)]{Fong2015}
{Fong}, W., {Berger}, E., {Margutti}, R., \& {Zauderer}, B.~A. 2015, \apj, 815,
  102

\bibitem[{Gao} {et~al.}(2013)]{GaoH2013}
{Gao}, H., {Ding}, X., {Wu}, X.-F., {Zhang}, B., \& {Dai}, Z.-G. 2013, \apj,
  771, 86

\bibitem[{Gao} {et~al.}(2016)]{Gao2016}
{Gao}, H., {Zhang}, B., \& {L{\"u}}, H.-J. 2016, \prd, 93, 044065

\bibitem[{Gao} \& {Fan}(2006)]{GaoWH2006}
{Gao}, W.-H., \& {Fan}, Y.-Z. 2006, \cjaa, 6, 513

\bibitem[{Gao} {et~al.}(2022)]{Gao2022}
{Gao}, Y., {Lai}, X.-Y., {Shao}, L., \& {Xu}, R.-X. 2022, \mnras, 509, 2758

\bibitem[{Gehrels} {et~al.}(2008)]{Gehrels2008}
{Gehrels}, N., {Barthelmy}, S.~D., {Burrows}, D.~N., {et~al.} 2008, \apj, 689,
  1161

\bibitem[{Guo} {et~al.}(2014)]{GLX2014}
{Guo}, Y.-J., {Lai}, X.-Y., \& {Xu}, R.-X. 2014, Chinese Physics C, 38, 055101

\bibitem[{Hartle}(1967)]{Hartle1967}
{Hartle}, J.~B. 1967, \apj, 150, 1005

\bibitem[{Hartle}(1973)]{Hartle1973}
{Hartle}, J.~B. 1973, \apss, 24, 385

\bibitem[{Hartle} \& {Thorne}(1968)]{Hartle1968}
{Hartle}, J.~B., \& {Thorne}, K.~S. 1968, \apj, 153, 807

\bibitem[{Hou} {et~al.}(2018)]{Hou2018}
{Hou}, S.-J., {Liu}, T., {Xu}, R.-X., {et~al.} 2018, \apj, 854, 104

\bibitem[{Hu} {et~al.}(2020)]{Hu2020}
{Hu}, H., {Kramer}, M., {Wex}, N., {Champion}, D.~J., \& {Kehl}, M.~S. 2020,
  \mnras, 497, 3118

\bibitem[{Kaghashvili}(1999)]{Kaghashvili1999}
{Kaghashvili}, E.~K. 1999, \grl, 26, 1817

\bibitem[Kaplan {et~al.}(2011)]{Kaplan2011}
Kaplan, D.~L., Kamble, A., van Kerkwijk, M.~H., \& Ho, W. C.~G. 2011,
  Astrophys. J., 736, 117

\bibitem[{Kisaka} {et~al.}(2017)]{Kisaka2017}
{Kisaka}, S., {Ioka}, K., \& {Sakamoto}, T. 2017, \apj, 846, 142

\bibitem[{Komatsu} {et~al.}(2009)]{Komatsu2009}
{Komatsu}, E., {Dunkley}, J., {Nolta}, M.~R., {et~al.} 2009, \apjs, 180, 330

\bibitem[{Lai} {et~al.}(2023{\natexlab{a}})]{LXX2023}
{Lai}, X., {Xia}, C., \& {Xu}, R. 2023{\natexlab{a}}, Advances in Physics X, 8,
  2137433

\bibitem[Lai {et~al.}(2013)]{LGX2013}
Lai, X.-Y., Gao, C.-Y., \& Xu, R.-X. 2013, Mon. Not. Roy. Astron. Soc., 431,
  3282

\bibitem[{Lai} {et~al.}(2023{\natexlab{b}})]{Lai2023MN}
{Lai}, X.~Y., {Wang}, W.~H., {Yuan}, J.~P., {et~al.} 2023{\natexlab{b}},
  \mnras, 523, 3967

\bibitem[{Lai} {et~al.}(2021)]{Lai2021RAA}
{Lai}, X.-Y., {Xia}, C.-J., {Yu}, Y.-W., \& {Xu}, R.-X. 2021, Research in
  Astronomy and Astrophysics, 21, 250

\bibitem[Lai \& Xu(2009{\natexlab{a}})]{LX2009a}
Lai, X.-Y., \& Xu, R.-X. 2009{\natexlab{a}}, Astropart. Phys., 31, 128

\bibitem[Lai \& Xu(2009{\natexlab{b}})]{LX2009b}
Lai, X.-Y., \& Xu, R.-X. 2009{\natexlab{b}}, Mon. Not. Roy. Astron. Soc., 398,
  L31

\bibitem[Lai {et~al.}(2018{\natexlab{a}})]{Lai2018RAA}
Lai, X.-Y., Yu, Y.-W., Zhou, E.-P., Li, Y.-Y., \& Xu, R.-X. 2018{\natexlab{a}},
  Res. Astron. Astrophys., 18, 024

\bibitem[Lai {et~al.}(2018{\natexlab{b}})]{Lai2018MN}
Lai, X.-Y., Yun, C.-A., Lu, J.-G., {et~al.} 2018{\natexlab{b}}, Mon. Not. Roy.
  Astron. Soc., 476, 3303

\bibitem[Lai {et~al.}(2019)]{LZX2019}
Lai, X.-Y., Zhou, E.-P., \& Xu, R.-X. 2019, Eur. Phys. J. A, 55, 60

\bibitem[{Li} {et~al.}(2016)]{LiA2016}
{Li}, A., {Zhang}, B., {Zhang}, N.-B., {et~al.} 2016, \prd, 94, 083010

\bibitem[{Lin} {et~al.}(2015)]{Lin2015}
{Lin}, M.-X., {Xu}, R.-X., \& {Zhang}, B. 2015, \apj, 799, 152

\bibitem[Lu {et~al.}(2019)]{Lu2019}
Lu, J.-G., Peng, B., Xu, R.-X., {et~al.} 2019, Sci. China-Phys. Mech. Astron.,
  62, 959505

\bibitem[{M{\'e}sz{\'a}ros} \& {Rees}(1997)]{Meszaros1997}
{M{\'e}sz{\'a}ros}, P., \& {Rees}, M.~J. 1997, \apjl, 482, L29

\bibitem[Michel(1988)]{Michel1988}
Michel, F. 1988, Phys. Rev. Lett., 60, 677

\bibitem[{Oppenheimer} \& {Volkoff}(1939)]{OV1939}
{Oppenheimer}, J.~R., \& {Volkoff}, G.~M. 1939, Physical Review, 55, 374

\bibitem[{Paczynski}(1991)]{Paczynski1991}
{Paczynski}, B. 1991, AcA, 41, 157

\bibitem[{Rowlinson} {et~al.}(2013)]{Rowlinson2013}
{Rowlinson}, A., {O'Brien}, P.~T., {Metzger}, B.~D., {Tanvir}, N.~R., \&
  {Levan}, A.~J. 2013, \mnras, 430, 1061

\bibitem[Ryan {et~al.}(2020)]{Ryan2020}
Ryan, G., van Eerten, H., Piro, L., \& Troja, E. 2020, \apj, 896, 166

\bibitem[{Sarin}(2021)]{Sarin2021PhDT}
{Sarin}, N. 2021, {The observational signatures of nascent neutron stars}, PhD
  thesis, Monash University, Australia

\bibitem[{Shapiro} \& {Teukolsky}(1983)]{Shapiro1983}
{Shapiro}, S.~L., \& {Teukolsky}, S.~A. 1983, {Black holes, white dwarfs and
  neutron stars. The physics of compact objects}

\bibitem[{Strang} \& {Melatos}(2019)]{Strang2019}
{Strang}, L.~C., \& {Melatos}, A. 2019, \mnras, 487, 5010

\bibitem[{Strang} {et~al.}(2021)]{Strang2021}
{Strang}, L.~C., {Melatos}, A., {Sarin}, N., \& {Lasky}, P.~D. 2021, \mnras,
  507, 2843

\bibitem[{Stratta} {et~al.}(2018)]{Stratta2018}
{Stratta}, G., {Dainotti}, M.~G., {Dall'Osso}, S., {Hernandez}, X., \& {De
  Cesare}, G. 2018, \apj, 869, 155

\bibitem[{Tu} \& {Marsch}(1997)]{Tu1997}
{Tu}, C.~Y., \& {Marsch}, E. 1997, \solphys, 171, 363

\bibitem[Wang {et~al.}(2017)]{Wang2017APJ}
Wang, W.-Y., Lu, J.-G., Tong, H., {et~al.} 2017, Astrophys. J., 837, 81

\bibitem[{Xiao} \& {Dai}(2019)]{Xiao2019}
{Xiao}, D., \& {Dai}, Z.-G. 2019, \apj, 878, 62

\bibitem[{Xu} \& {Guo}(2017)]{XG2016}
{Xu}, R., \& {Guo}, Y. 2017, in Centennial of General Relativity: A
  Celebration, ed. C.~A. {Zen Vasconcellos}, 119

\bibitem[Xu(2003)]{Xu2003}
Xu, R.-X. 2003, Astrophys. J. Lett., 596, L59

\bibitem[Xu {et~al.}(1999)]{Xu1999ApJL}
Xu, R.-X., Qiao, G.-J., \& Zhang, B. 1999, Astrophys. J. Lett., 522, L109

\bibitem[Yu {et~al.}(2009)]{YuYW2009}
Yu, Y.-W., Cao, X.-F., \& Zheng, X.-P. 2009, \apj, 706, L221

\bibitem[{Zhang}(2013)]{ZhangB2013}
{Zhang}, B. 2013, \apjl, 763, L22

\bibitem[{Zhang} \& {M{\'e}sz{\'a}ros}(2001)]{Zhang2001}
{Zhang}, B., \& {M{\'e}sz{\'a}ros}, P. 2001, \apjl, 552, L35

\bibitem[{Zhang} \& {M{\'e}sz{\'a}ros}(2002)]{Zhang2002}
{Zhang}, B., \& {M{\'e}sz{\'a}ros}, P. 2002, \apj, 566, 712

\bibitem[Zhou {et~al.}(2004)]{Zhou2004}
Zhou, A.-Z., Xu, R.-X., Wu, X.-J., \& Wang, N. 2004, Astropart. Phys., 22, 73

\bibitem[Zhou {et~al.}(2014)]{Zhou2014}
Zhou, E.-P., Lu, J.-G., Tong, H., \& Xu, R.-X. 2014, Mon. Not. Roy. Astron.
  Soc., 443, 2705

\bibitem[{Zhou} {et~al.}(2023)]{Zhou2023}
{Zhou}, E., {Gao}, Y., {Zhou}, Y., {et~al.} 2023, arXiv e-prints,
  arXiv:2305.10682

\end{thebibliography}

\label{lastpage}

\end{document}